\documentclass[aip,amsmath,amssymb,reprint]{revtex4-1}
\usepackage{epsfig}
\usepackage{amsmath}
\graphicspath{{figures/}}
\usepackage{epsfig}
\usepackage{amsmath}
\usepackage{color}
\usepackage{graphicx,epsfig}
\usepackage{color}
\usepackage[latin9]{inputenc}
\usepackage{float}
\usepackage{natbib}

\begin{document}

\title{Anomalous Behavior in the Nucleation of Ice at Negative Pressures}

\author{V. Bianco}
\author{P. Montero de Hijes}
\author{C. P. Lamas}
\author{E. Sanz}
\author{C. Vega$^{*}$}
\affiliation{$^1$Departamento de Quimica Fisica, 
Facultad de Quimica, Universidad Complutense de 
Madrid, Ciudad Universitaria, Madrid 28040, Spain}

%
\begin{abstract}

Ice nucleation is a phenomenon that, despite the relevant implications for life, 
atmospheric sciences and technological applications, is far from being completely understood, 
especially under extreme thermodynamic conditions. 
In this work we present a computational investigation of the homogeneous 
ice nucleation at negative pressures. By means of the {\it seeding technique} we 
estimate the size of the ice critical nucleus $N_c$ for the TIP4P/Ice water model. This is done 
 along the isotherms $230\;{\rm K}$, $240\;{\rm K}$, and $250\;{\rm K}$, from positive to negative pressures 
until reaching the liquid-gas kinetic stability limit (where cavitation cannot be avoided). 
We find that $N_c$ is non-monotonic upon depressurization, reaching a minimum at 
negative pressures in the doubly metastable region of water. 
According to classical nucleation theory we establish the nucleation 
rate $J$ and the surface tension $\gamma$, revealing a 
retracing behavior of both when the liquid-gas kinetic stability limit is approached. 
We also predict a reentrant behavior of the homogeneous nucleation line.
 The re-entrance of these properties is related to 
 the re-entrance of the coexistence line at negative pressure, revealing new anomalies of water. 
The results of this work suggest the possibility of having metastable samples of liquid water 
for long times at negative pressure provided that heterogeneous nucleation is suppressed.

\end{abstract}

\maketitle
$^*$Corresponding author: cvega@quim.ucm.es

Ice formation is possibly the most important liquid-to-solid 
 transition, 
being relevant in 
cryo-biology, food storage, material science,
 and Earth science \cite{Gettelman2010, Coluzza2017,Nitzbon2020,JohnMorris2013,BarDolev2016,Broekaert2011,Russo2014a,Bintanja2013}. 
Homogeneous nucleation is the mechanism through which thermal fluctuations in a 
pure liquid below coexistence 
  induce the formation of crystal nuclei 
that, when sufficiently large (critical size), trigger the crystallization.

At ambient pressure and a few kelvins below 
 coexistence, 
 the size of the critical nucleus $N_c$ is huge and the probability of forming 
spontaneously in pure supercooled water is negligible \cite{Espinosa2016, Niu2019}. 
Consequently, in nature, ice is formed essentially 
via heterogeneous nucleation \cite{Sanz2013a}. 

Several investigations have addressed the behavior of supercooled liquid water 
at negative pressure 
\cite{experimentos_science,Azouzi2013,stanley_negative_pressure, Henderson1987, Q1991, Imre2002, Caupin2015,gallo2016,pablo_D}, 
with little attention paid to ice nucleation
 \cite{Marcolli2017, KannoScience1975}. Here, 
we fill this gap by exploring the 
 homogeneous ice nucleation from positive to negative
 pressure $P$ at constant temperature $T$. 

By means of molecular dynamic simulations of the TIP4P/Ice 
water model \cite{Abascal2005}
 -- probably the best atomistic model to study ice properties and with a 
well known phase diagram\cite{Abascal2005,science_2020}-- 
we reveal that the isothermal variation of 
 $N_c$ is non-monotonic. 
For any isothermal path, a minimum is always observed at $P<0$. 
  This retracing behavior 
is linked to the re-entrance of the coexistence line. 

All the simulations have been performed using GROMACS \cite{Hess2008}, adopting: 
  i) a time step of $2\;{\rm fs}$; 
 ii) the Noose-Hoover thermostat with a relaxation time of $1\;{\rm ps}$; 
iii) the Parrinello-Rahman barostat with a relaxation time of $2\;{\rm ps}$; 
 iv) the particle-mesh-Ewald algorithm of order 4, with Fourier spacing 
of $0.1\;{\rm nm}$ to solve the electrostatic interaction; 
  v) a cutoff of $0.9\;{\rm nm}$ both for the Lennard-Jones and Coulomb interactions; 
 vi) long range corrections to the Lennard-Jones interaction.

\begin{figure}
	\begin{center}
\includegraphics[width=0.35\textwidth]{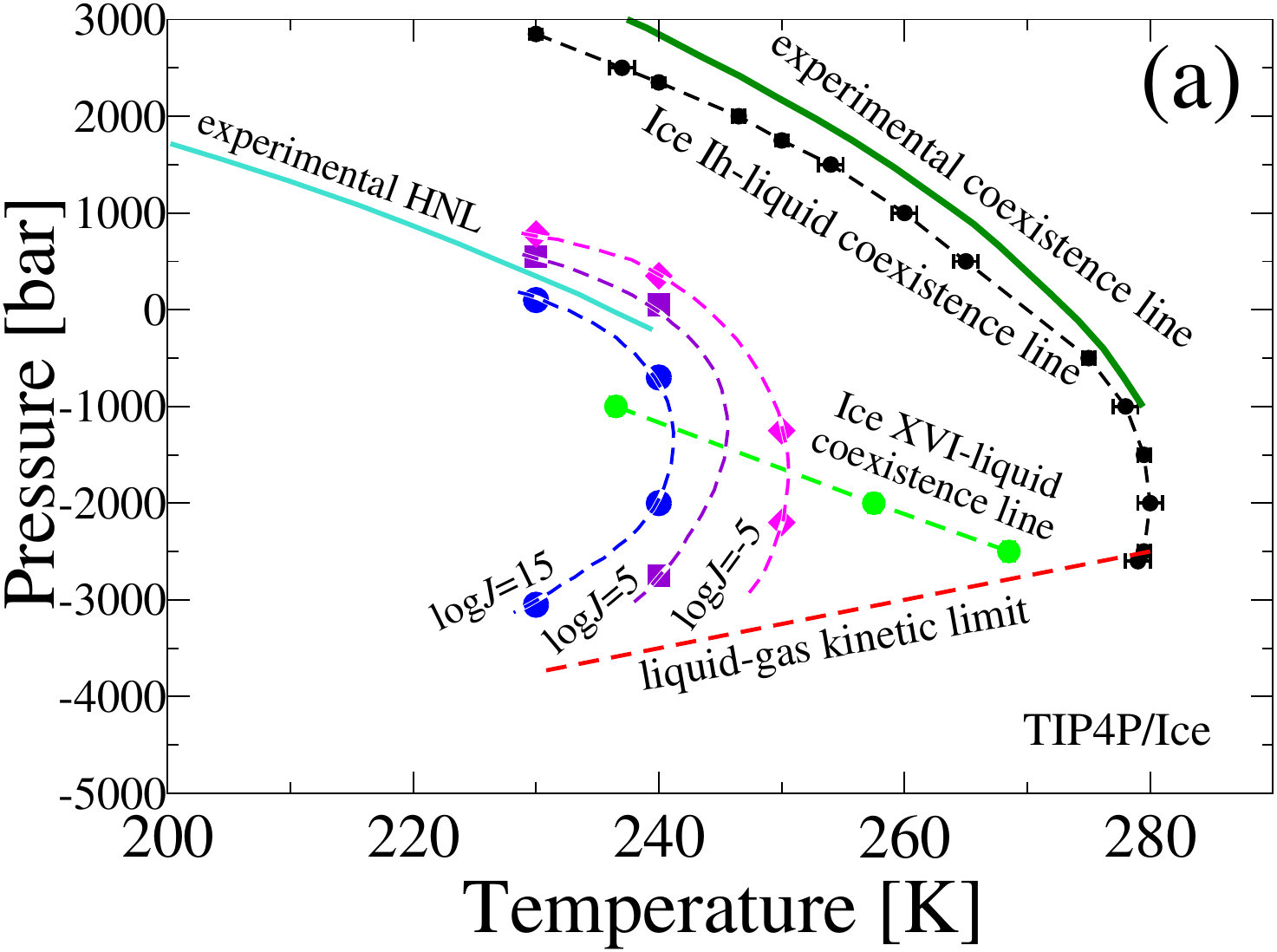}
\includegraphics[width=0.35\textwidth]{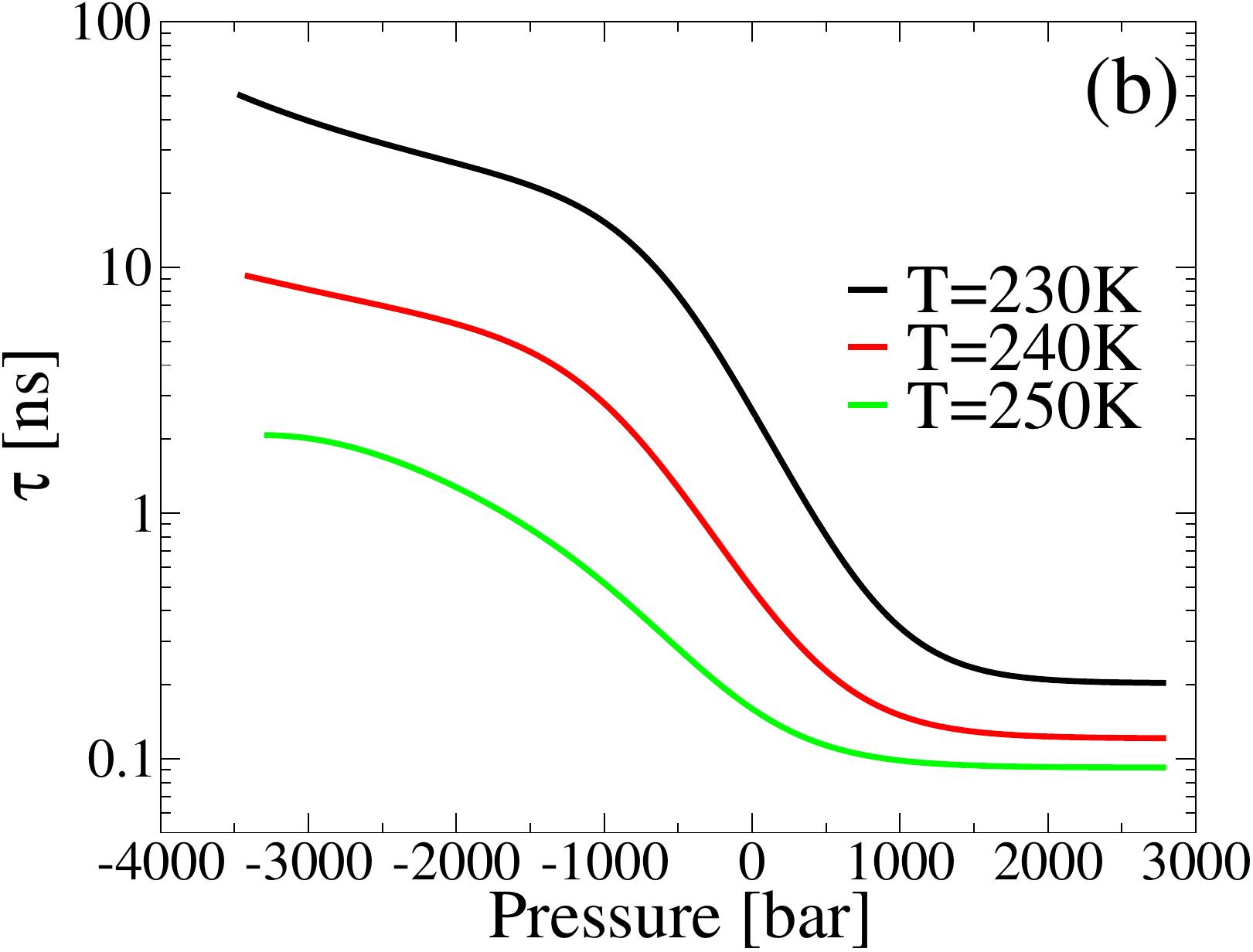}
\caption{
(a) Phase diagram in the $T-P$ plane of the TIP4P/Ice water model showing:
i) the ice Ih-liquid coexistence line (black points);  
ii) the liquid-gas kinetic limit (red dashed line); 
iii) the iso-lines with constant logarithm of the nucleation rate: $\log_{10} J /( m^{-3} s^{-1} ) =15, 5, -5$ (see Fig. \ref{fig:gamma_logJ}b). 
iv) the fitting of experimental homogeneous nucleation line (HNL) \cite{KannoScience1975} 
and ice Ih-liquid coexistence proposed by Marcolli \cite{Marcolli2017};
 v) the ice XVI-liquid coexistence curve (green points).
The HNL (where the formation of ice can not be avoided) from experiments corresponds approximately to an iso-line 
$\log_{10} J /( m^{-3} s^{-1} ) =15$ \cite{Espinosa2016}. 
 (b) Relaxation time $\tau \equiv (0.31\text{ nm})^{2}/(6D)$ to diffuse the square of the diameter 
of a water molecule as a function of $P$.}
\label{fig:coexistence_line}
	\end{center}
	\end{figure}

\begin{figure*}
        \begin{center}

	\includegraphics[scale=0.39]{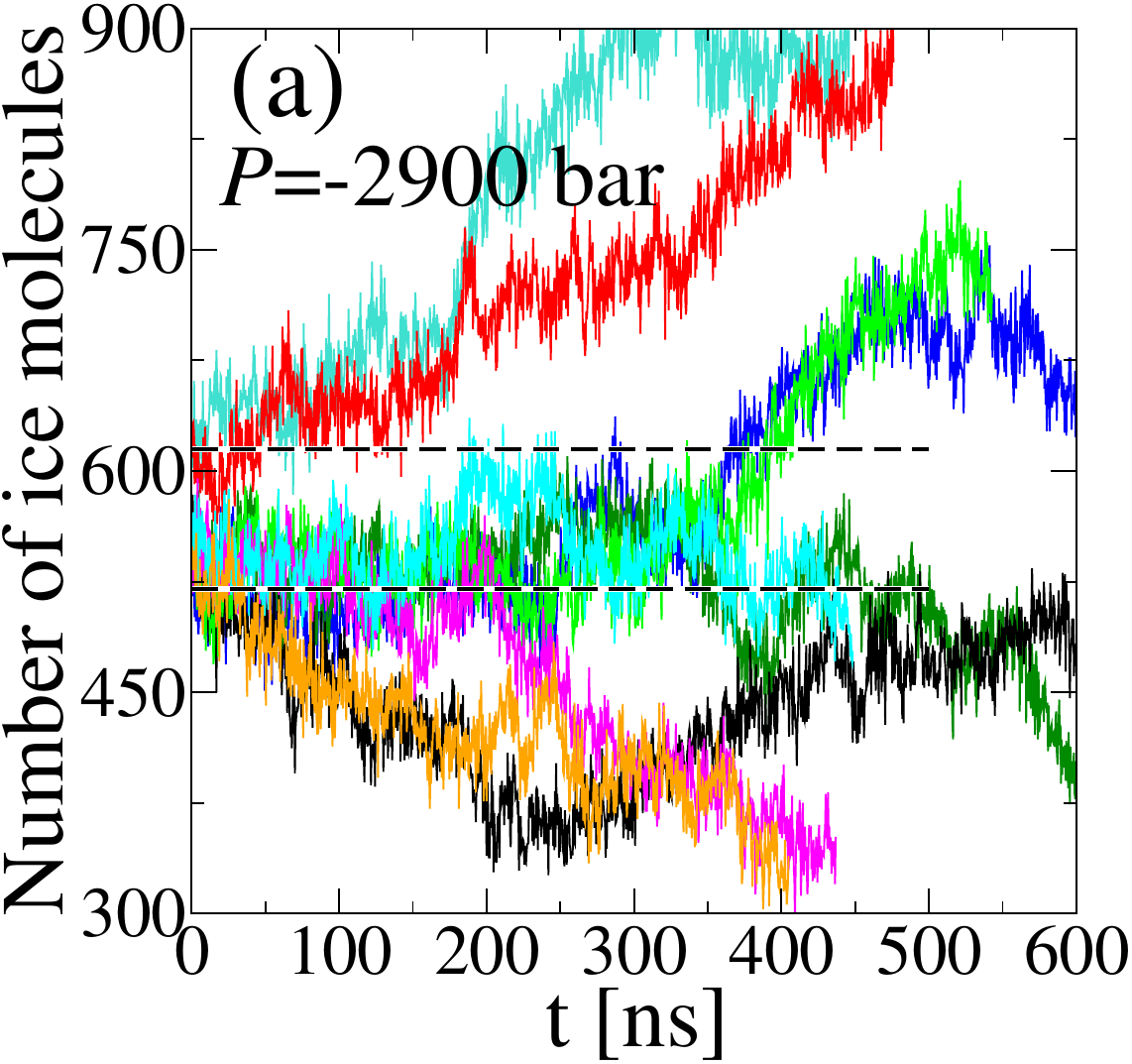}
\includegraphics[scale=0.39]{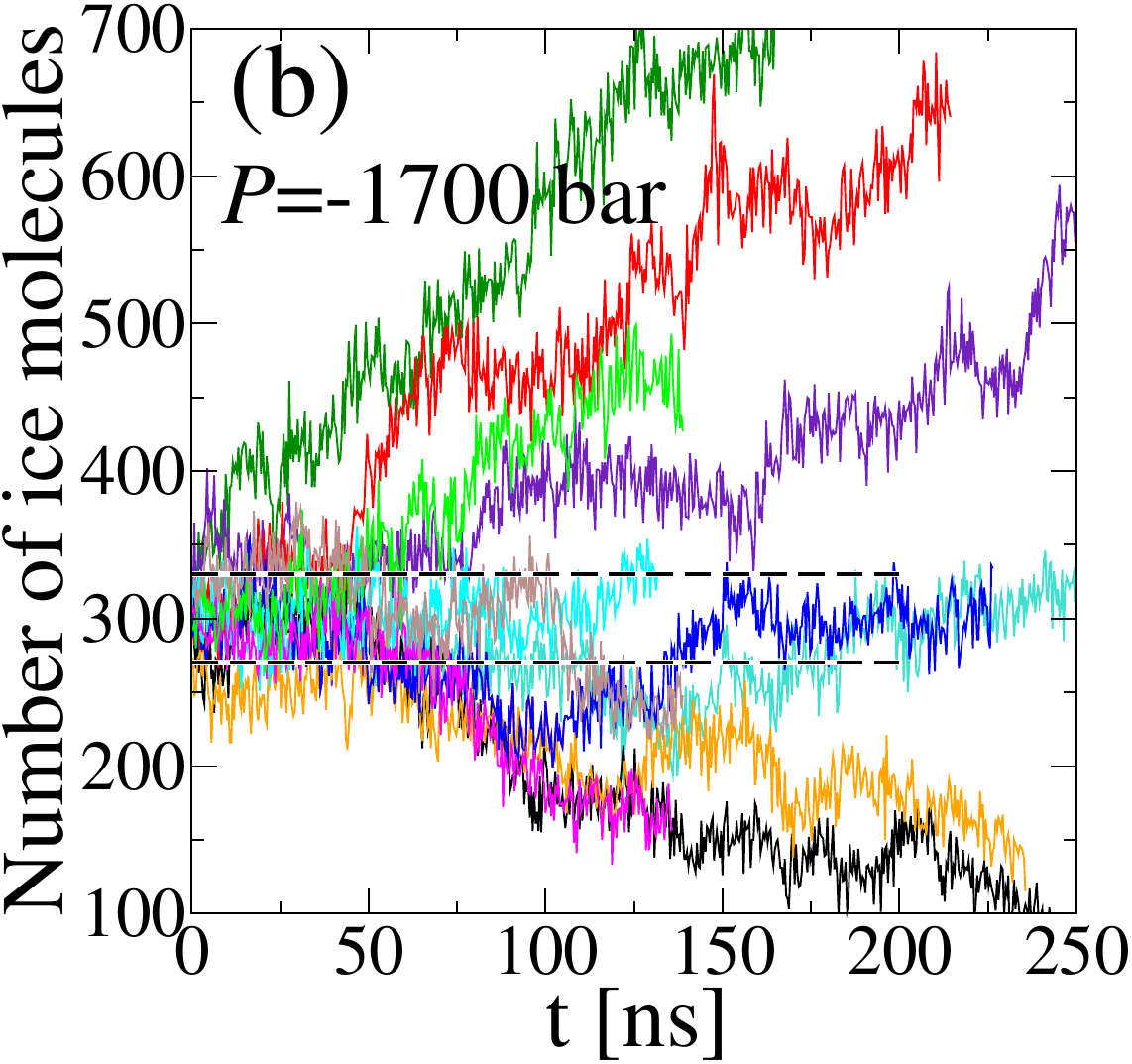}
\includegraphics[scale=0.39]{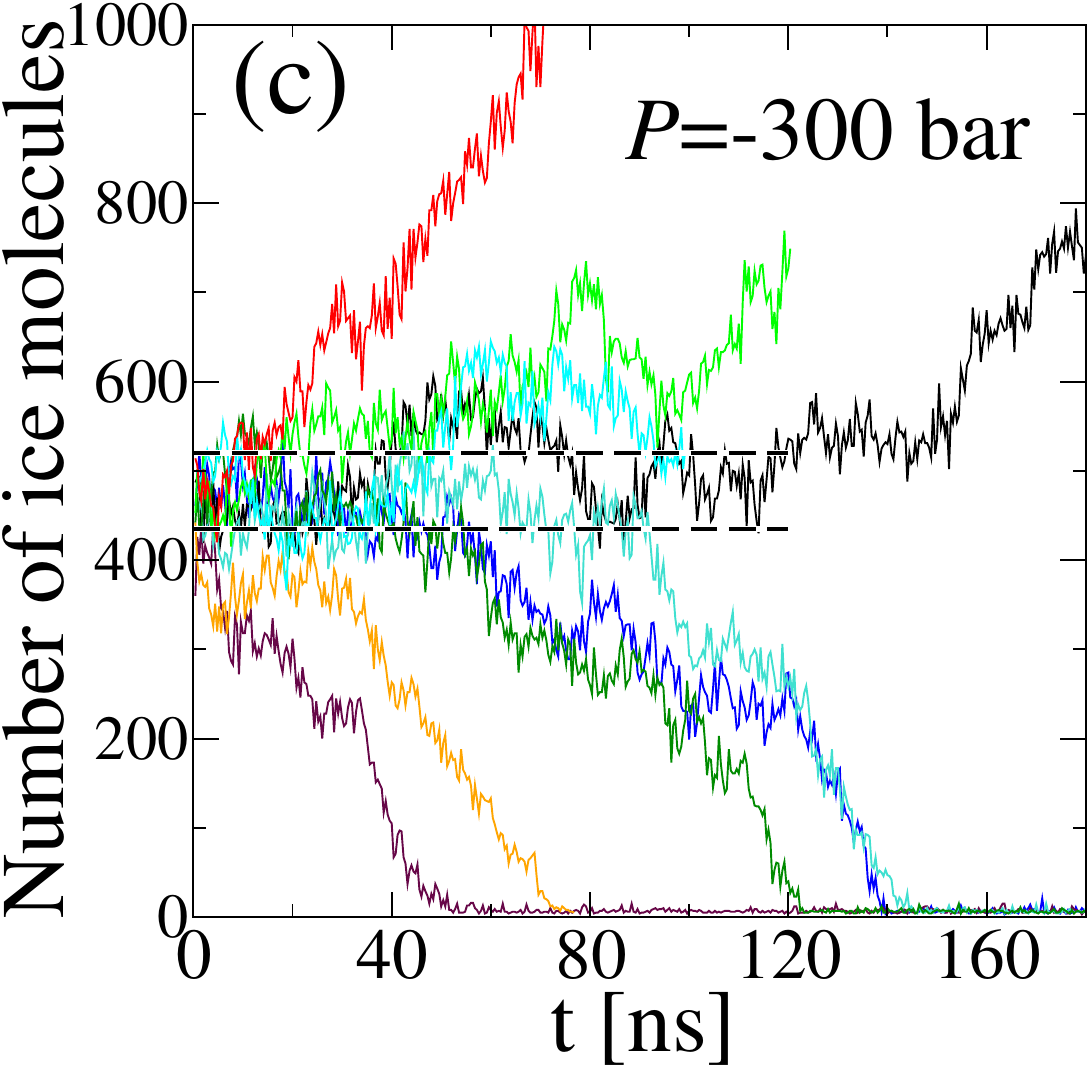}
	\hspace{0.2cm} \includegraphics[scale=0.39]{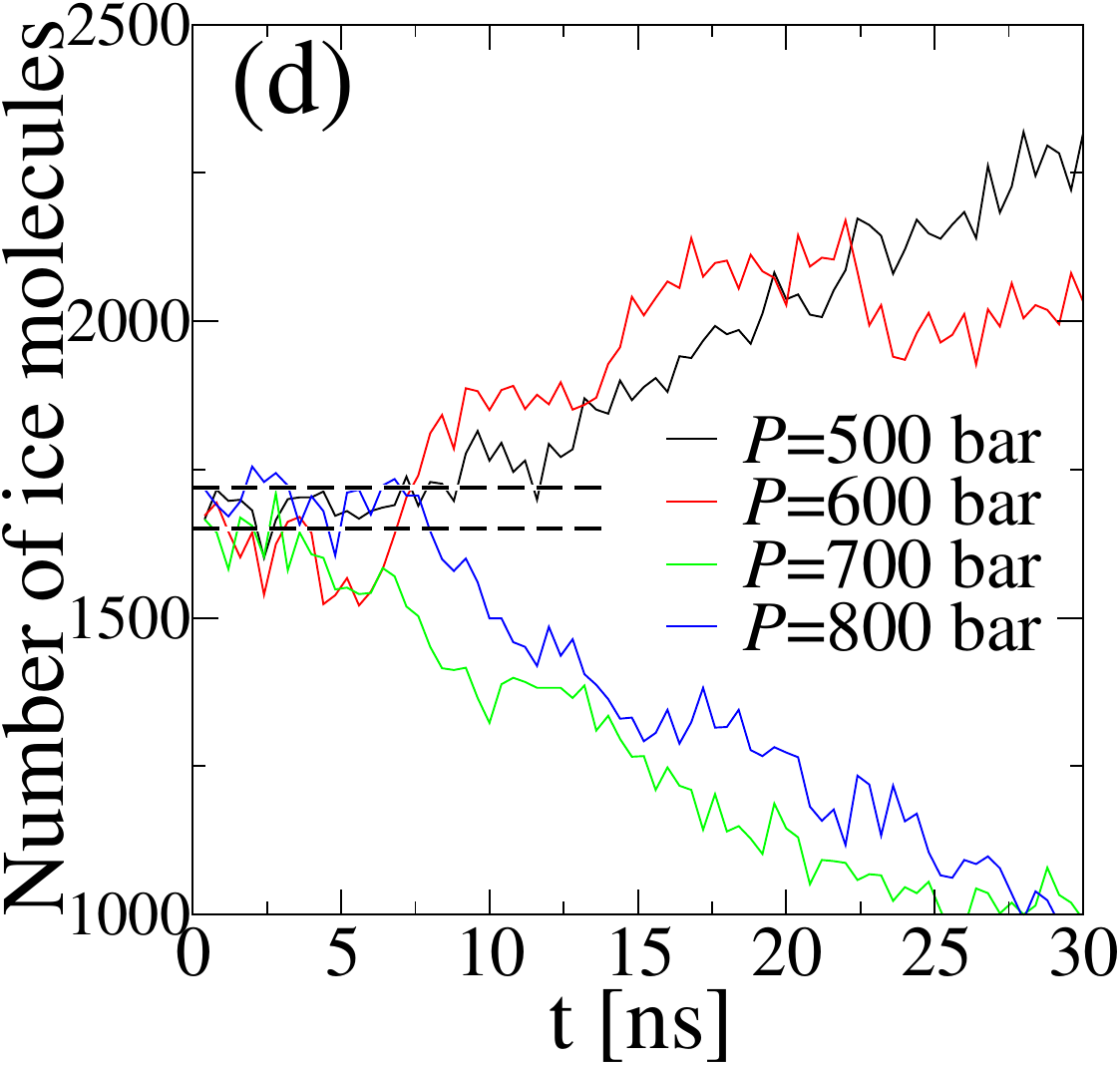}
\caption{Time evolution of the ice seed along the isotherm $T=240\;{\rm K}$ at: 
(a) $P=-2900\;{\rm bar}$; (b) $P=-1700\;{\rm bar}$; (c) $P=-300\;{\rm bar}$; (d) different positive pressures. 
Dashed lines mark the boundaries of the estimated critical size $N_c$ (established as the difference between the size of the largest seed melting and the smallest seed growing).
In a-c $P$ was constant and we changed the size of the initial ice cluster. In d), 
the initial size of the cluster was identical but it was studied at different $P$. From the results of d) one estimates that the cluster is critical at $P\sim 650\,{\rm bar}$.}
\label{fig:seeding}
        \end{center}
	\end{figure*}

Given the polymorphism of ice, 
we investigate the most stable structure. 
  According to Matsui et al. \cite{Matsui}, Ih is clearly the most stable ice up to 
extreme negative pressures, 
where ice Ih and the recently discovered ice XVI \cite{icexvi} are close in terms of stability. 
Hence, first we determine the solid-liquid coexistence lines for ice Ih 
and ice XVI for the TIP4P/Ice model via direct coexistence simulations
[see the Supporting Material (SM)\cite{SM_ref}].  
The $T$-$P$ loci of the coexistence lines are shown in Fig. \ref{fig:coexistence_line}a.
For $P>-3000\;{\rm bar}$ the ice XVI coexistence line is lower than the Ih coexistence line, meaning that the Ih structure is more stable than the ice XVI for this pressure range. 
The triple point water-Ih-ice XVI is estimated at $P\sim -3000\;{\rm bar}$ 
and  $T\sim 278\;{\rm K}$, similar to what has been found for the TIP4P/2005 water model 
 \cite{virtual}. Therefore, we have conducted our investigation with ice Ih, although for $P<-3000\;{\rm bar}$ the nucleation of ice XVI could be relevant
.

 We then estimate the liquid-gas kinetic stability limit 
defined as the $T-P$ locus where 2000 liquid water  molecules were stable for at least  
$100\;{\rm ns}$.   
   For any thermodynamic state 
in this work 
 it is  possible to reach a metastable equilibrium, 
as the relaxation time 
is smaller than $40\;{\rm ns}$ 
(Fig. \ref{fig:coexistence_line}b).
As can be seen, before this stability limit is reached, there is a re-entrance 
of the coexistence line with a turning point  
$dT/dP=0$ at $P\sim-2000\;{\rm bar}$ and $T\sim 280\;{\rm K}$. 
In 1982, Speedy conjectured a re-entering behavior of the liquid-gas spinodal
 \cite{Speedy1982}. 
This has been recently demonstrated to be the case for
colloidal systems \cite{Rovigatti2017} but ruled out for water \cite{De03} as further seen in this work. Nevertheless, Henderson and Speedy later conjectured 
a reentrance in the coexistence line \cite{Henderson1987} 
(also suggested by Bridgman in 1912 \cite{bridgman}). Speedy's estimation of the turning point
of the coexistence line occurring at $P\sim-1750\, {\rm bar}$ and $T\sim 283\, {\rm K}$ \cite{Henderson1987} is surprisingly close to our numerical finding.

Although experimentally inaccessible for large (in absolute value) negative pressures, 
the possible re-entrance 
 of the solid-liquid coexistence 
line affects the kinetic and thermodynamic properties of the accessible metastable region 
of the supercooled water's phase diagram at negative pressure. 
This phenomenon resembles what has been largely discussed in the last decades
 about the origin of water's anomalies
 \cite{speedy76,llcp, Kim:2009aa, Stokely2010, Russo2014a, Bianco2014, Palmer2014, Pallares2014,nilsson_2015, gallo2016, Perakis2017, Bianco2019, Gallo2019}. 
There, the possible presence of a 
second liquid-liquid critical point in a low $T$-high $P$ region of the phase 
diagram (confirmed recently for the TIP4P/Ice model\cite{science_2020} although 
experimentally prohibitive) would determine the increase of the fluctuations 
(and related thermodynamic response functions) 
in the  accessible metastable region of the phase diagram.

To evaluate 
 $N_c$ we follow the {\it seeding } computational approach, 
introduced by Bai and Li \cite{Bai2005}, and widely adopted in nucleation studies 
\cite{Sanz2013a, Espinosa2014, Espinosa2016a, Espinosa2016b, Espinosa2018, 
Lifanov2016, Leoni2019, montero2019,MonterodeHijes2020, montero2020young,
tipeev, Knott2012, pereyra2011, Dasgupta2020, Niu2020,zimmermann2015} 
(see SM for details). This scheme consists in introducing a spherical crystal seed 
of a given size into a bulk of supercooled liquid and let it evolve 
at constant $T$ and $P$. Seeds whose size is larger than $N_c(T,P)$ 
grow spanning the entire system, while those below $N_c(T,P)$  melt. When the size is critical, 
the crystal seed can grow or melt with equal probability. Hence, by comparing the time 
evolution of the size of the crystal seeds
at certain $T$ and $P$, it is possible to establish  $N_c$ (within a certain resolution). 
One can either study seeds differing in size at fixed state 
point $T-P$ (Fig. \ref{fig:seeding}a-c) or a given 
seed at different $T-P$  (Fig. \ref{fig:seeding}d).

\begin{figure}
\begin{center}
\includegraphics[width=0.23\textwidth]{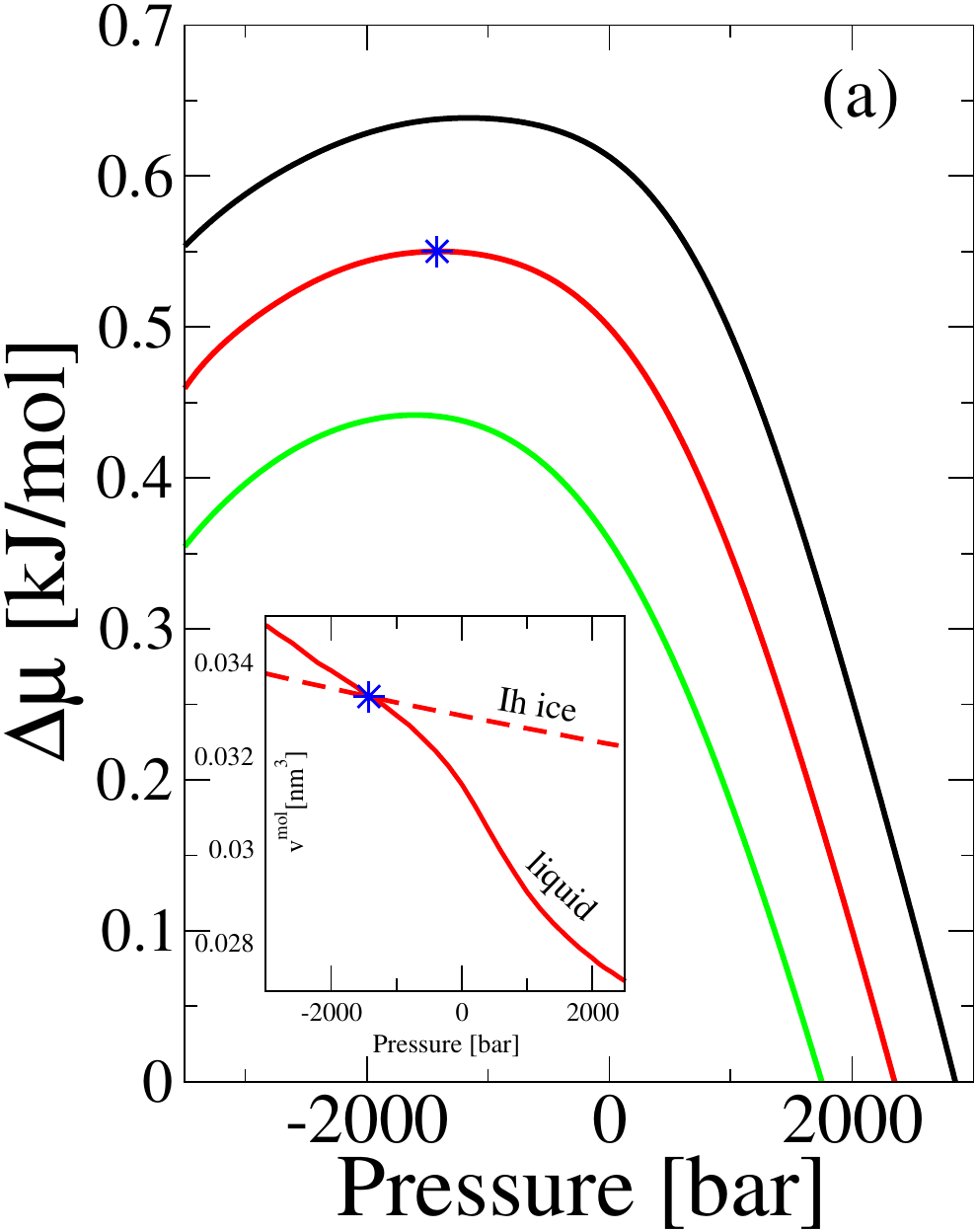}
\includegraphics[width=0.21\textwidth]{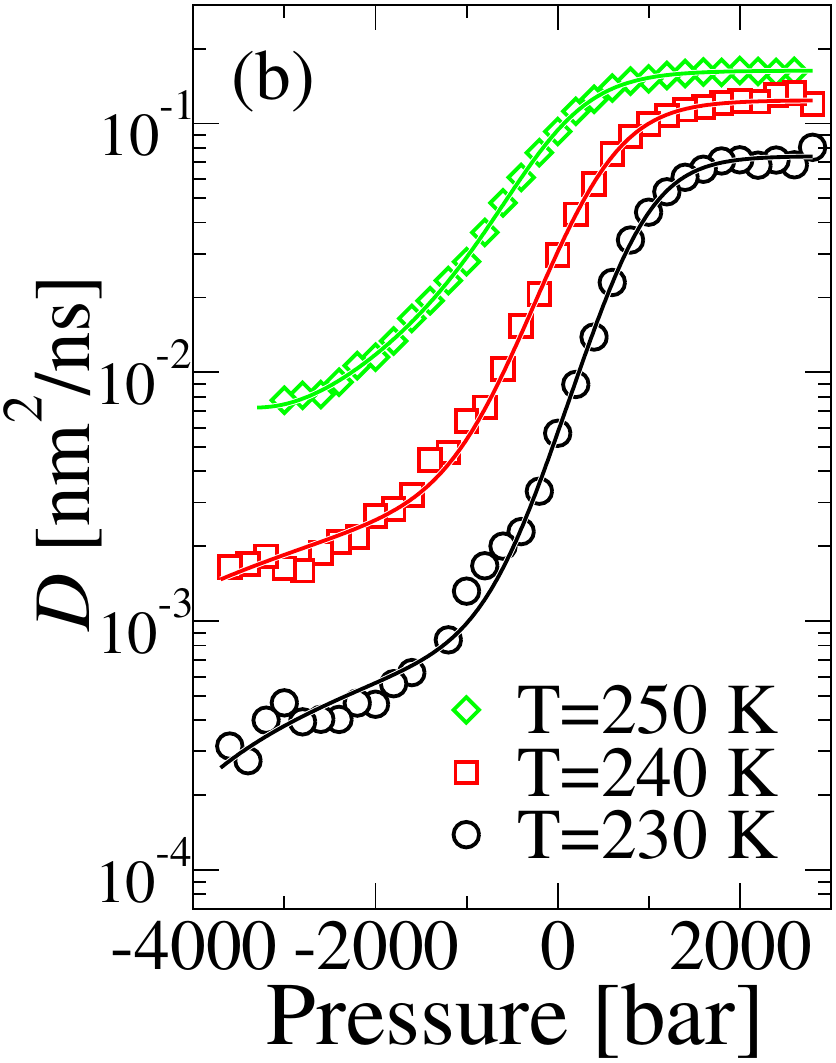}\vspace{0.3cm}
\includegraphics[width=0.33\textwidth]{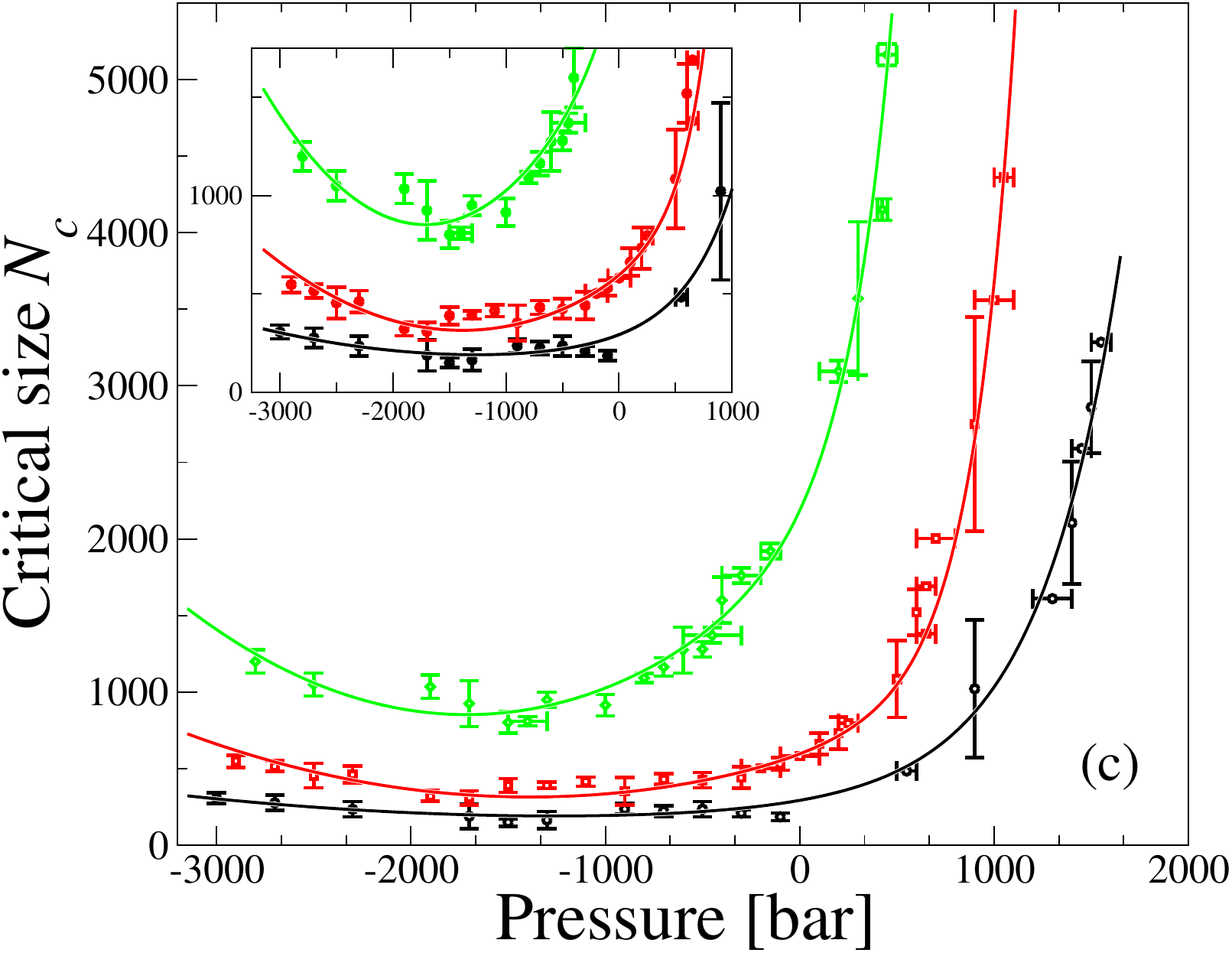}
\includegraphics[width=0.37\textwidth]{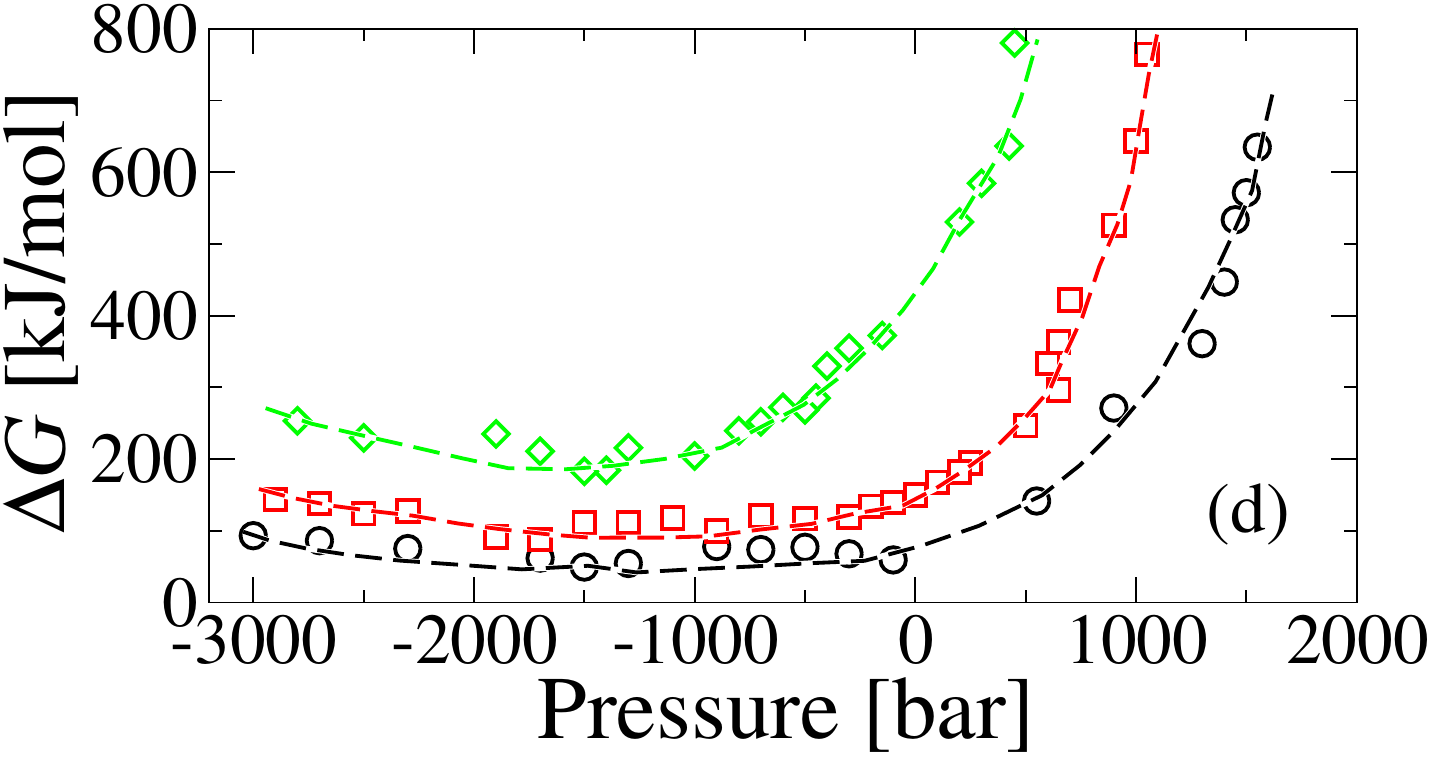}
\caption{Different observables as a function of $P$ along the isotherms $T=230\;{\rm K}$, $T=240\;{\rm K}$, and $T=250\;{\rm K}$ (legend for all panels in (b)):
	(a) Difference in chemical potential $\Delta \mu$. Volume per molecule of Ih ice and
	liquid water at $T=240\;{\rm K}$ in inset. Blue star indicators are at the same $P$.
(b) Diffusion coefficient $D$. Lines are fitting curves reported in the SM. 
(c) Size of the ice critical nucleus $N_c$. Inset zooms on the region where the minima are observed. Lines are fitting functions given in the SM.
	(d) Gibbs free energy barrier $\Delta G$ (zoom in at the minima including error bars in SM). 
	Dashed lines are guides for the eye. }
\label{fig:delta_mu_D}
\end{center}
\end{figure}

\begin{figure}[t!]
        \begin{center}

	\includegraphics[width=0.35\textwidth]{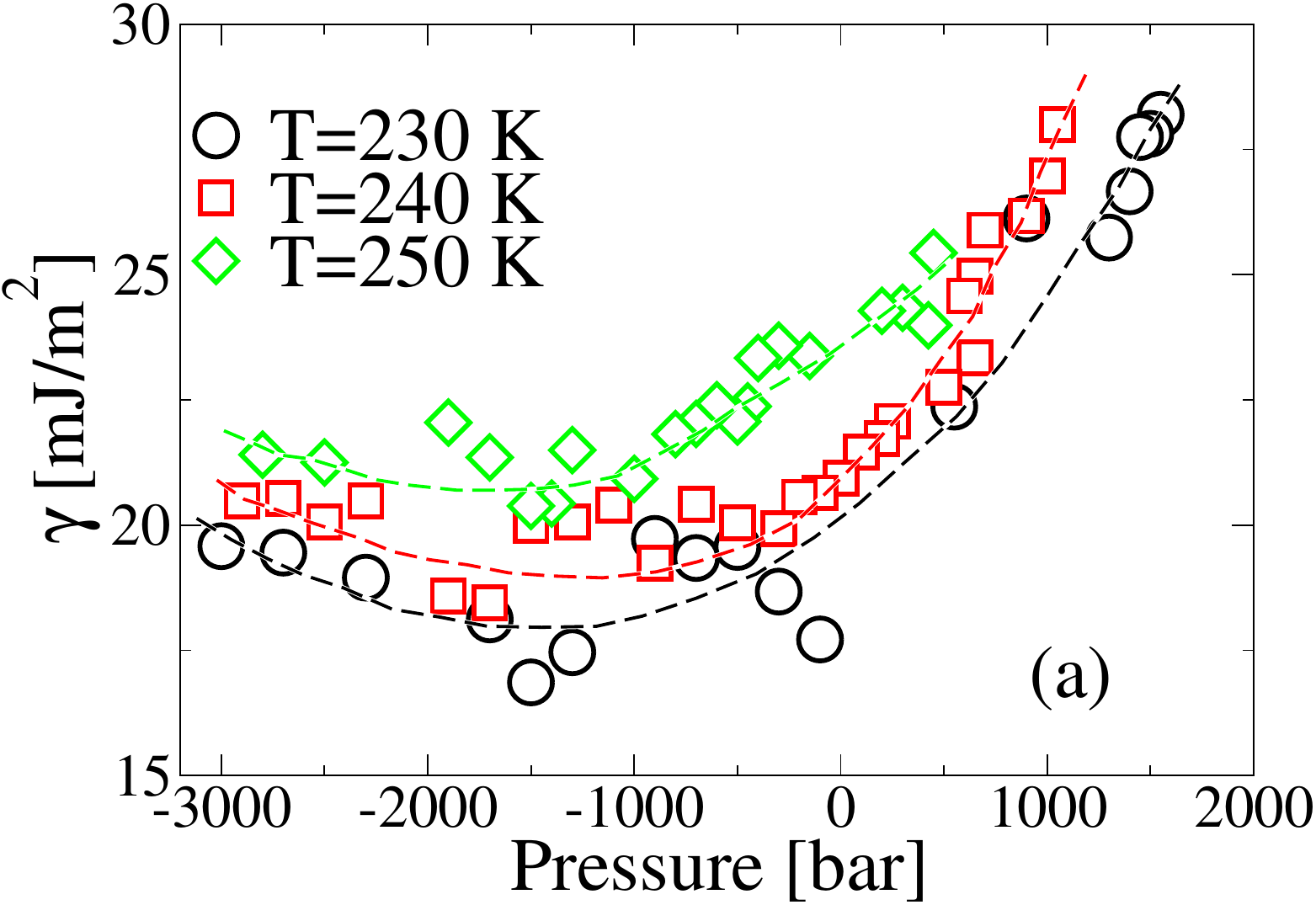}
\includegraphics[width=0.35\textwidth]{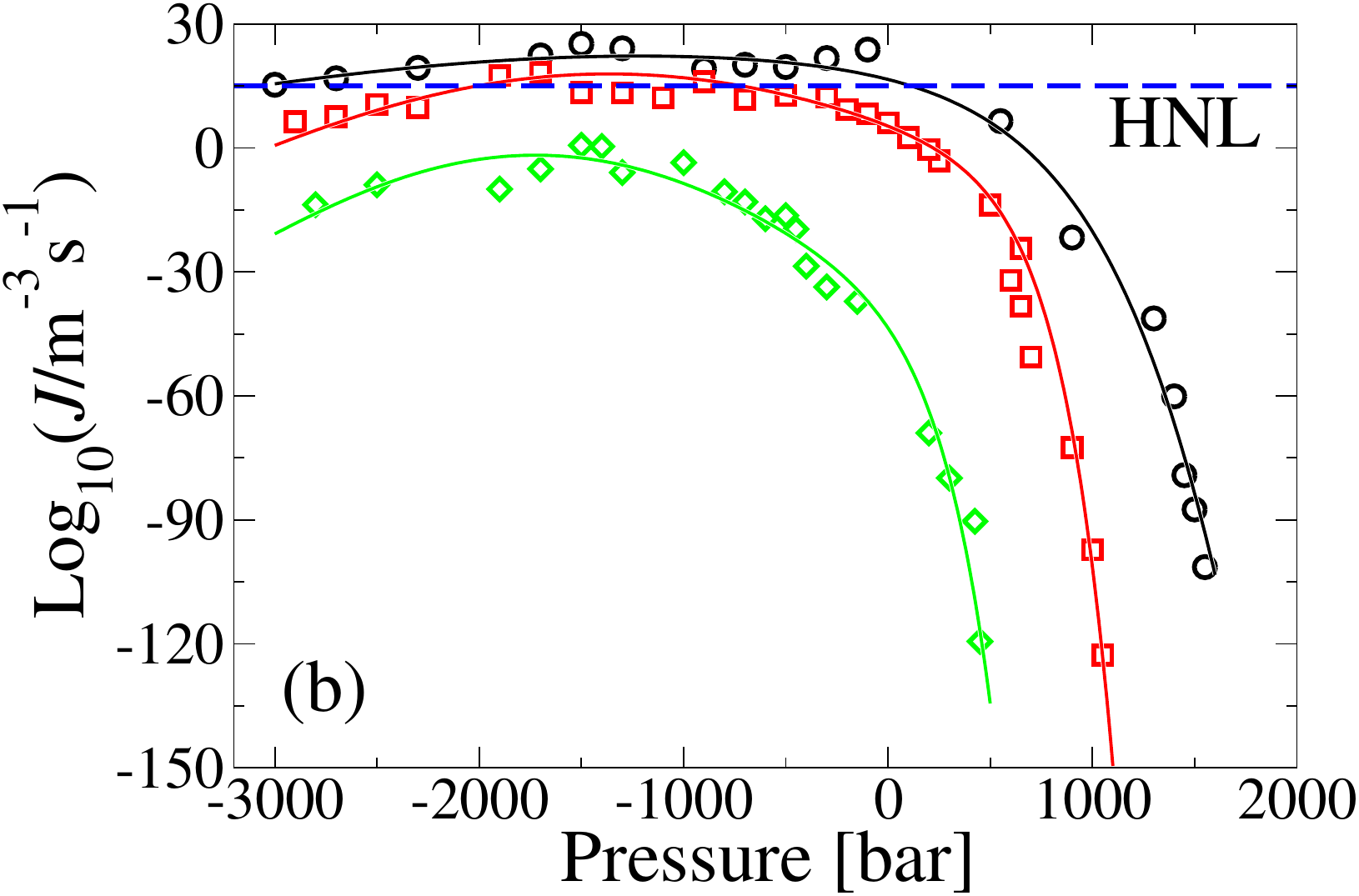}
\caption{ Isothermal variations (at $230\;{\rm K}$, $240\;{\rm K}$, and $250\;{\rm K}$) with $P$ of: 
(a) Surface tension $\gamma$. Dashed lines are guides for the eye. 
	(b) The logarithm of $J$ (zoom in at the maxima including error bars in SM). Same legend than (a).
	Dashed blue line marks the homogeneous nucleation line (HNL) given by $\log_{10}  J /$( m$^{-3}$ s$^{-1}$ )  $= 15$. 
Continues lines are analytical expression according to Eq. (\ref{eq:J}), using the fitting curves $\rho_{\rm liq}(T,P)$ shown in Fig 2 of the SM and 
and  $D(T,P)$ and $N_c(T,P)$ shown in Fig. \ref{fig:delta_mu_D}b and \ref{fig:delta_mu_D}c respectively,  all reported in the SM. 
By estimating the $(T,P)$ state points where $\log_{10}  J /$( m$^{-3}$ s$^{-1}$ ) $= 15, 5, -5$ we have drawn the iso-$J$ lines shown in Fig. \ref{fig:coexistence_line}. 
	}
\label{fig:gamma_logJ}
        \end{center}
	\end{figure}

We investigate the $T=(230, 240, 250)\;{\rm K}$ isotherms with $N\sim 46000$ water molecules. For any $T$ we explore $P\in[-3500:2400]\;{\rm bar}$, i.e.
 from the liquid-gas kinetic stability limit at negative pressures up
 to the melting point at positive pressures. 
The number of ice molecules is determined according to the Lechner-Dellago order parameter $\bar{q}_6$ \cite{Lechner2008}, with cutoff distance of $3.5\, \AA$.
Molecules (within the cutoff distance) above the threshold $\bar{q}_{6,t}$ are labeled as ice,
 whereas those with smaller values are labeled as liquid. 
The value of $\bar{q}_{6,t}$ which depends on $P$ and $T$ is determined according to the mislabeling criterion\cite{Espinosa2016a} (see the SM for details). 
In Fig. \ref{fig:seeding} we show the time  evolution of the number of ice particles 
in the crystal seed
along the isotherm 
 $T=240\;{\rm K}$, with runs spanning in some cases up to the $\mu{\rm s}$.
 On average any $N_c$ has been identified by means of $\sim10$ independent runs.
All the estimated $N_c$ along the three isotherms are presented in Fig. \ref{fig:delta_mu_D}c 
and reported in the SM.
As shown in Fig. \ref{fig:delta_mu_D}c, 
from positive pressures, $N_c$ largely decreases upon decreasing $P$, reaching a quasi-constant value at negative $P$. 
By further decreasing $P$ we find that $N_c$ increases again. 
In all cases the minimum value of $N_c$ is observed at $P<0$.  
  In particular, we find that the minimum value of $N_c$ 
is
$\sim 150$ for $T=230\;{\rm K}$, 
$\sim 310$ for $T=240\;{\rm K}$, and 
$\sim 800$ for $T=250\;{\rm K}$.  
From $N_c(T,P)$ we can extract the $P$-dependence of the surface tension $\gamma(T,P)$ 
following the classical nucleation theory (CNT). 
Indeed, according to CNT, $\gamma$ can be expressed as \cite{Espinosa2016b}

\begin{equation}
 \gamma(T,P) = \left [ \dfrac{3\rho_{\rm ice}(T,P)^2 |\Delta \mu (T,P)|^3 N_c(T,P)}{32\pi} \right ]^{1/3} ,
 \label{eq:gamma}
\end{equation}

being $\rho_{\rm ice}(T,P)$ the ice density at $(T,P)$, and $\Delta\mu (T,P) \equiv \mu_{\rm liq}(T,P) - \mu_{\rm ice}(T,P)$ the difference in chemical 
potential between  water ($\mu_{\rm liq}$) and ice ($\mu_{\rm ice}$). 
As discussed in the SM (see Fig. 3 and Table II), $\Delta \mu$ can be computed by 
thermodynamic integration. 
In Fig. \ref{fig:delta_mu_D}a we report $\Delta \mu$ along the isotherms of interest, 
showing a retracing behavior whose maximum value is always reached at $P<0$. 
The maxima of $\Delta\mu$ are observed at $(T=230\: {\rm K},P\sim -1160\: {\rm bar})$, $(T=240\: {\rm K},P\sim -1405\: {\rm bar})$,
 and $(T=250\: {\rm K},P\sim -1605\: {\rm bar})$, 
roughly coinciding with the thermodynamic state points where $N_c(P)$ exhibits a minimum. 
The presence of the maximum of $\Delta \mu$ is due to the crossing of ice and liquid water 
densities along the corresponding isotherm  as shown in Fig. \ref{fig:delta_mu_D}a 
inset (see also Fig. 3(c-e) of the SM). 
From Eq. \ref{eq:gamma} we obtain $\gamma$ which is shown in Fig. \ref{fig:gamma_logJ}a. 
As can be seen, $\gamma$ decreases upon decreasing $P$, reaching a quasi-constant value at negative $P$, 
then increasing again at largely (in absolute value) negative $P$.
We can extract the values of $\gamma$ at $P=1\: {\rm bar}$: 
$\gamma\sim 20\: {\rm mJ/m^2}$ for  $T=230\;{\rm K}$, 
$\gamma\sim 21\: {\rm mJ/m^2}$ for  $T=240\;{\rm K}$, and 
$\gamma\sim 24\: {\rm mJ/m^2}$ for  $T=250\;{\rm K}$. 
These values are in agreement with the fitting expression for $\gamma(T,P=1\: {\rm bar})$ 
reported in Ref. \cite{Espinosa2016b,montero2019} 
($\gamma =$ 19.1, 21.8, $24.5\;{\rm mJ/m}^2$ respectively).

Following Becker and D\"oring \cite{Becker1935} we can express the nucleation rate $J(T,P)$ as
\begin{equation}
 J(T,P) = \rho_{\rm liq} f^+ Z \exp (-\Delta G_c/k_BT)\qquad,
 \label{eq:J}
\end{equation}
where $\rho_{\rm liq}$ is the density of the liquid phase, 
and $\Delta G_c=N_c|\Delta\mu(T,P)|/2$
is the free energy barrier associated to the formation of a critical cluster.
 Also, $k_B$ is the Boltzmann constant and $Z \equiv \sqrt{|\Delta\mu|/(6\pi k_B T N_c )}$ 
is the Zeldovich factor which depends on the curvature of $\Delta G$ around the barrier top and 
is related to the width of the critical region. 
The attachment rate $f^+$ can be approximated as $f^+ = 24DN_c^{2/3}/\lambda^2$,  
where $D$ is the diffusion coefficient of the supercooled water (shown in Fig. \ref{fig:delta_mu_D}b) and $\lambda=3.8\AA$ according to previous work \cite{Espinosa2014}. 
Such an approximation succesfully works at positive pressure \cite{Espinosa2014}. 
We test it at negative pressure by computing $f^{+}$ rigorously as in Ref. \cite{auer2001} 
from $\langle (N(t) - N(0))^{2}\rangle = 2f^+t $,  
 at $T=240\;{\rm K}$ for $P = -300,-1700,-2900\;{\rm bar}$ finding good agreement, as reported in SM
(the deviations do not affect the resulting value of $J$).

The  $P$-dependence of $\Delta G_c$ is shown in Fig. \ref{fig:delta_mu_D}d.
Then, we estimate $J$ via Eq. \ref{eq:J}. In Fig. \ref{fig:gamma_logJ}b, the returning of $J$ at negative $P$ is revealed. 
The maxima of $J$ are found at $P<0$ for all the explored $T$ 
and mark the $(T,P)$ conditions where homogeneous nucleation occurs more easily 
as recently suggested by Marcolli \cite{Marcolli2017}. In this particular case, 
$J\sim10^{22}\;{\rm m}^{-3}{\rm s}^{-1}$ for ($T=230\;{\rm K}$, $P\sim -1200\;{\rm bar}$), 
$J\sim10^{18}\;{\rm m}^{-3}{\rm s}^{-1}$ for ($T=240\;{\rm K}$, $P\sim -1350\;{\rm bar}$), and
$J\sim10^{-1}\;{\rm m}^{-3}{\rm s}^{-1}$ for ($T=250\;{\rm K}$, $P\sim -1750\;{\rm bar}$). 
Accordingly, the estimated surviving time $\tau$ of a water droplet with volume $V_{\rm ice}\sim 5$ cm$^3$ characteristic of the inclusion experiments at negative $P$ \cite{Azouzi2013, Caupin2015}
 is $\tau\equiv (J V_{\rm ice})^{-1} \sim 2\times10^{-17}\;{\rm s}$ at $T=230\;{\rm  K}$, $\tau \sim2\times10^{-13}\;{\rm s}$ at  $T=240\;{\rm K}$, and  $\tau \sim2\times10^{6}\;{\rm s}$ 
at $T=250\;{\rm K}$. 
Thus, this work suggests that $245\;{\rm K}$ - $250\;{\rm K}$ is the minimum 
 $T$  at which one can study macroscopic 
samples of water at $P<0$ without freezing the entire sample in a few seconds.


Finally, we compute the iso-nucleation rate lines for TIP4P/Ice (representing the loci of points in the $P-T$ plane 
where the nucleation rates have identical values). They are presented in 
Fig. \ref{fig:coexistence_line} for $\log_{10} J /$( m$^{-3}$ s$^{-1}$ )$ =15,5,-5$. 
As  can be seen, the iso-nucleation 
rate lines  present reentrant behavior. The curve $\log_{10} J /$( m$^{-3}$ s$^{-1}$ )$ =15$ is of particular interest 
as it can be regarded as an estimate of the 
experimental homogeneous nucleation line (HNL) \cite{Espinosa2016} where freezing of droplets of a few microns occurs in a few seconds and cannot 
be avoided, thus, representing the solid-liquid limit of stability. 
In the $230\;{\rm K}$ isotherm, $\log_{10} J /$( m$^{-3}$ s$^{-1}$ )$ =15$ occurs at $P=1\;{\rm bar}$ (see also Fig. \ref{fig:coexistence_line}). 
The results of this work suggest that the HNL presents reentrant behavior, 
 another anomaly of water arising at the confluence of 
low $T$ and negative $P$\cite{altabet2017,holten2017} that, to the best of our knowledge, 
has not been reported before.

In conclusion, we use computer simulations of the TIP4P/Ice water model  to estimate 
the size of the critical nucleus $N_c$ along 
three isotherms $T=230\: {\rm K}$, $T=240\: {\rm K}$, and $T=250\: {\rm K}$. 
We cover from typical positive coexistence pressures 
 until approaching the liquid-gas kinetic stability curve (under 100 ns of observation). 
We show how $N_c$ does not change monotonically 
exhibiting a minimum at negative $P$ and increasing again in the 
vicinity of the  stability limit. Accordingly, the nucleation rate $J$ and the surface 
tension $\gamma$ show a retracing behavior, with a maximum of $J$ and a minimum of $\gamma$ 
both occurring at negative $P$. 
Our findings reveal new water anomalies as the retracing behavior of 
$N_c$, $\gamma$ and $J$ along the isotherms when going from positive to negative pressures. 
We also predict anomalous behavior of the homogeneous nucleation line, 
which again presents reentrant behavior at negative pressures. 
This can be regarded as a smoking gun of the re-entrance of the melting curve 
which can be evaluated experimentally only up to moderate values of negative pressures \cite{Henderson1987, Q1991, Imre2002, Azouzi2013, Caupin2015}.

The authors acknowledge the project PID2019-105898GB-C21 of the MEC. V. B. acknowledges the support from the European Commission
 through the Marie Sk\l odowska-Curie Fellowship No. 748170 ProFrost. P.M.d.H. acknowledges the financial support from the FPI grant no. BES-2017-080074. 
C.P.L thanks MEC for a predoctoral FPU grant FPU18/03326 and also Ayuntamiento
        de Madrid for a Residencia de Estudiantes grant. 
The authors acknowledge the computer resources and technical assistance provided by the RES
	 and the Vienna Scientific Cluster (VSC).


\begin{thebibliography}{65}
\expandafter\ifx\csname natexlab\endcsname\relax\def\natexlab#1{#1}\fi
\expandafter\ifx\csname bibnamefont\endcsname\relax
  \def\bibnamefont#1{#1}\fi
\expandafter\ifx\csname bibfnamefont\endcsname\relax
  \def\bibfnamefont#1{#1}\fi
\expandafter\ifx\csname citenamefont\endcsname\relax
  \def\citenamefont#1{#1}\fi
\expandafter\ifx\csname url\endcsname\relax
  \def\url#1{\texttt{#1}}\fi
\expandafter\ifx\csname urlprefix\endcsname\relax\def\urlprefix{URL }\fi
\providecommand{\bibinfo}[2]{#2}
\providecommand{\eprint}[2][]{\url{#2}}

\bibitem[{\citenamefont{Gettelman et~al.}(2010)\citenamefont{Gettelman, Liu,
  Ghan, Morrison, Park, Conley, Klein, Boyle, Mitchell, and
  Li}}]{Gettelman2010}
\bibinfo{author}{\bibfnamefont{A.}~\bibnamefont{Gettelman}},
  \bibinfo{author}{\bibfnamefont{X.}~\bibnamefont{Liu}},
  \bibinfo{author}{\bibfnamefont{S.~J.} \bibnamefont{Ghan}},
  \bibinfo{author}{\bibfnamefont{H.}~\bibnamefont{Morrison}},
  \bibinfo{author}{\bibfnamefont{S.}~\bibnamefont{Park}},
  \bibinfo{author}{\bibfnamefont{A.~J.} \bibnamefont{Conley}},
  \bibinfo{author}{\bibfnamefont{S.~A.} \bibnamefont{Klein}},
  \bibinfo{author}{\bibfnamefont{J.}~\bibnamefont{Boyle}},
  \bibinfo{author}{\bibfnamefont{D.~L.} \bibnamefont{Mitchell}},
  \bibnamefont{and} \bibinfo{author}{\bibfnamefont{J.-L.~F.} \bibnamefont{Li}},
  \bibinfo{journal}{Journal of Geophysical Research}
  \textbf{\bibinfo{volume}{115}}, \bibinfo{pages}{D18216}
  (\bibinfo{year}{2010}).

\bibitem[{\citenamefont{Coluzza et~al.}(2017)\citenamefont{Coluzza, Creamean,
  Rossi, Wex, Alpert, Bianco, Boose, Dellago, Felgitsch,
  Fr{\"{o}}hlich-Nowoisky et~al.}}]{Coluzza2017}
\bibinfo{author}{\bibfnamefont{I.}~\bibnamefont{Coluzza}},
  \bibinfo{author}{\bibfnamefont{J.}~\bibnamefont{Creamean}},
  \bibinfo{author}{\bibfnamefont{M.}~\bibnamefont{Rossi}},
  \bibinfo{author}{\bibfnamefont{H.}~\bibnamefont{Wex}},
  \bibinfo{author}{\bibfnamefont{P.}~\bibnamefont{Alpert}},
  \bibinfo{author}{\bibfnamefont{V.}~\bibnamefont{Bianco}},
  \bibinfo{author}{\bibfnamefont{Y.}~\bibnamefont{Boose}},
  \bibinfo{author}{\bibfnamefont{C.}~\bibnamefont{Dellago}},
  \bibinfo{author}{\bibfnamefont{L.}~\bibnamefont{Felgitsch}},
  \bibinfo{author}{\bibfnamefont{J.}~\bibnamefont{Fr{\"{o}}hlich-Nowoisky}},
  \bibnamefont{et~al.}, \bibinfo{journal}{Atmosphere}
  \textbf{\bibinfo{volume}{8}}, \bibinfo{pages}{138} (\bibinfo{year}{2017}).

\bibitem[{\citenamefont{Nitzbon et~al.}(2020)\citenamefont{Nitzbon, Westermann,
  Langer, Martin, Strauss, Laboor, and Boike}}]{Nitzbon2020}
\bibinfo{author}{\bibfnamefont{J.}~\bibnamefont{Nitzbon}},
  \bibinfo{author}{\bibfnamefont{S.}~\bibnamefont{Westermann}},
  \bibinfo{author}{\bibfnamefont{M.}~\bibnamefont{Langer}},
  \bibinfo{author}{\bibfnamefont{L.~C.~P.} \bibnamefont{Martin}},
  \bibinfo{author}{\bibfnamefont{J.}~\bibnamefont{Strauss}},
  \bibinfo{author}{\bibfnamefont{S.}~\bibnamefont{Laboor}}, \bibnamefont{and}
  \bibinfo{author}{\bibfnamefont{J.}~\bibnamefont{Boike}},
  \bibinfo{journal}{Nature Communications} \textbf{\bibinfo{volume}{11}},
  \bibinfo{pages}{2201} (\bibinfo{year}{2020}).

\bibitem[{\citenamefont{{John Morris} and Acton}(2013)}]{JohnMorris2013}
\bibinfo{author}{\bibfnamefont{G.}~\bibnamefont{{John Morris}}}
  \bibnamefont{and} \bibinfo{author}{\bibfnamefont{E.}~\bibnamefont{Acton}},
  \bibinfo{journal}{Cryobiology} \textbf{\bibinfo{volume}{66}},
  \bibinfo{pages}{85} (\bibinfo{year}{2013}).

\bibitem[{\citenamefont{{Bar Dolev} et~al.}(2016)\citenamefont{{Bar Dolev},
  Braslavsky, and Davies}}]{BarDolev2016}
\bibinfo{author}{\bibfnamefont{M.}~\bibnamefont{{Bar Dolev}}},
  \bibinfo{author}{\bibfnamefont{I.}~\bibnamefont{Braslavsky}},
  \bibnamefont{and} \bibinfo{author}{\bibfnamefont{P.~L.}
  \bibnamefont{Davies}}, \bibinfo{journal}{Annual Review of Biochemistry}
  \textbf{\bibinfo{volume}{85}}, \bibinfo{pages}{515} (\bibinfo{year}{2016}).

\bibitem[{\citenamefont{Broekaert et~al.}(2011)\citenamefont{Broekaert,
  Heyndrickx, Herman, Devlieghere, and Vlaemynck}}]{Broekaert2011}
\bibinfo{author}{\bibfnamefont{K.}~\bibnamefont{Broekaert}},
  \bibinfo{author}{\bibfnamefont{M.}~\bibnamefont{Heyndrickx}},
  \bibinfo{author}{\bibfnamefont{L.}~\bibnamefont{Herman}},
  \bibinfo{author}{\bibfnamefont{F.}~\bibnamefont{Devlieghere}},
  \bibnamefont{and}
  \bibinfo{author}{\bibfnamefont{G.}~\bibnamefont{Vlaemynck}},
  \bibinfo{journal}{Food Microbiology} \textbf{\bibinfo{volume}{28}},
  \bibinfo{pages}{1162} (\bibinfo{year}{2011}).

\bibitem[{\citenamefont{Russo et~al.}(2014)\citenamefont{Russo, Romano, and
  Tanaka}}]{Russo2014a}
\bibinfo{author}{\bibfnamefont{J.}~\bibnamefont{Russo}},
  \bibinfo{author}{\bibfnamefont{F.}~\bibnamefont{Romano}}, \bibnamefont{and}
  \bibinfo{author}{\bibfnamefont{H.}~\bibnamefont{Tanaka}},
  \bibinfo{journal}{Nature Materials} \textbf{\bibinfo{volume}{13}},
  \bibinfo{pages}{733} (\bibinfo{year}{2014}).

\bibitem[{\citenamefont{Bintanja et~al.}(2013)\citenamefont{Bintanja, van
  Oldenborgh, Drijfhout, Wouters, and Katsman}}]{Bintanja2013}
\bibinfo{author}{\bibfnamefont{R.}~\bibnamefont{Bintanja}},
  \bibinfo{author}{\bibfnamefont{G.~J.} \bibnamefont{van Oldenborgh}},
  \bibinfo{author}{\bibfnamefont{S.~S.} \bibnamefont{Drijfhout}},
  \bibinfo{author}{\bibfnamefont{B.}~\bibnamefont{Wouters}}, \bibnamefont{and}
  \bibinfo{author}{\bibfnamefont{C.~A.} \bibnamefont{Katsman}},
  \bibinfo{journal}{Nature Geoscience} \textbf{\bibinfo{volume}{6}},
  \bibinfo{pages}{376} (\bibinfo{year}{2013}).

\bibitem[{\citenamefont{Espinosa
  et~al.}(2016{\natexlab{a}})\citenamefont{Espinosa, Zaragoza, Rosales-Pelaez,
  Navarro, Valeriani, Vega, and Sanz}}]{Espinosa2016}
\bibinfo{author}{\bibfnamefont{J.~R.} \bibnamefont{Espinosa}},
  \bibinfo{author}{\bibfnamefont{A.}~\bibnamefont{Zaragoza}},
  \bibinfo{author}{\bibfnamefont{P.}~\bibnamefont{Rosales-Pelaez}},
  \bibinfo{author}{\bibfnamefont{C.}~\bibnamefont{Navarro}},
  \bibinfo{author}{\bibfnamefont{C.}~\bibnamefont{Valeriani}},
  \bibinfo{author}{\bibfnamefont{C.}~\bibnamefont{Vega}}, \bibnamefont{and}
  \bibinfo{author}{\bibfnamefont{E.}~\bibnamefont{Sanz}},
  \bibinfo{journal}{Physical Review Letters} \textbf{\bibinfo{volume}{117}},
  \bibinfo{pages}{135702} (\bibinfo{year}{2016}{\natexlab{a}}).

\bibitem[{\citenamefont{Niu et~al.}(2019)\citenamefont{Niu, Yang, and
  Parrinello}}]{Niu2019}
\bibinfo{author}{\bibfnamefont{H.}~\bibnamefont{Niu}},
  \bibinfo{author}{\bibfnamefont{Y.~I.} \bibnamefont{Yang}}, \bibnamefont{and}
  \bibinfo{author}{\bibfnamefont{M.}~\bibnamefont{Parrinello}},
  \bibinfo{journal}{Physical Review Letters} \textbf{\bibinfo{volume}{122}},
  \bibinfo{pages}{245501} (\bibinfo{year}{2019}).

\bibitem[{\citenamefont{Sanz et~al.}(2013)\citenamefont{Sanz, Vega, Espinosa,
  Caballero-Bernal, Abascal, and Valeriani}}]{Sanz2013a}
\bibinfo{author}{\bibfnamefont{E.}~\bibnamefont{Sanz}},
  \bibinfo{author}{\bibfnamefont{C.}~\bibnamefont{Vega}},
  \bibinfo{author}{\bibfnamefont{J.~R.} \bibnamefont{Espinosa}},
  \bibinfo{author}{\bibfnamefont{R.}~\bibnamefont{Caballero-Bernal}},
  \bibinfo{author}{\bibfnamefont{J.~L.~F.} \bibnamefont{Abascal}},
  \bibnamefont{and}
  \bibinfo{author}{\bibfnamefont{C.}~\bibnamefont{Valeriani}},
  \bibinfo{journal}{Journal of the American Chemical Society}
  \textbf{\bibinfo{volume}{135}}, \bibinfo{pages}{15008}
  (\bibinfo{year}{2013}).

\bibitem[{\citenamefont{Green et~al.}(1990)\citenamefont{Green, Durden, Wolf,
  and Angell}}]{experimentos_science}
\bibinfo{author}{\bibfnamefont{J.}~\bibnamefont{Green}},
  \bibinfo{author}{\bibfnamefont{D.}~\bibnamefont{Durden}},
  \bibinfo{author}{\bibfnamefont{G.}~\bibnamefont{Wolf}}, \bibnamefont{and}
  \bibinfo{author}{\bibfnamefont{C.}~\bibnamefont{Angell}},
  \bibinfo{journal}{Science} \textbf{\bibinfo{volume}{249}},
  \bibinfo{pages}{649} (\bibinfo{year}{1990}).

\bibitem[{\citenamefont{Azouzi et~al.}(2013)\citenamefont{Azouzi, Ramboz,
  Lenain, and Caupin}}]{Azouzi2013}
\bibinfo{author}{\bibfnamefont{M.~E.~M.} \bibnamefont{Azouzi}},
  \bibinfo{author}{\bibfnamefont{C.}~\bibnamefont{Ramboz}},
  \bibinfo{author}{\bibfnamefont{J.-F.} \bibnamefont{Lenain}},
  \bibnamefont{and} \bibinfo{author}{\bibfnamefont{F.}~\bibnamefont{Caupin}},
  \bibinfo{journal}{Nature Physics} \textbf{\bibinfo{volume}{9}},
  \bibinfo{pages}{38} (\bibinfo{year}{2013}).

\bibitem[{\citenamefont{Netz et~al.}(2001)\citenamefont{Netz, Starr, Stanley,
  and Barbosa}}]{stanley_negative_pressure}
\bibinfo{author}{\bibfnamefont{P.}~\bibnamefont{Netz}},
  \bibinfo{author}{\bibfnamefont{F.}~\bibnamefont{Starr}},
  \bibinfo{author}{\bibfnamefont{H.}~\bibnamefont{Stanley}}, \bibnamefont{and}
  \bibinfo{author}{\bibfnamefont{M.}~\bibnamefont{Barbosa}},
  \bibinfo{journal}{J. Chem. Phys.} \textbf{\bibinfo{volume}{115}},
  \bibinfo{pages}{344} (\bibinfo{year}{2001}).

\bibitem[{\citenamefont{Henderson and Speedy}(1987)}]{Henderson1987}
\bibinfo{author}{\bibfnamefont{S.~J.} \bibnamefont{Henderson}}
  \bibnamefont{and} \bibinfo{author}{\bibfnamefont{R.~J.}
  \bibnamefont{Speedy}}, \bibinfo{journal}{The Journal of Physical Chemistry}
  \textbf{\bibinfo{volume}{91}}, \bibinfo{pages}{3069} (\bibinfo{year}{1987}).

\bibitem[{\citenamefont{Zheng et~al.}(1991)\citenamefont{Zheng, Durben, Wolf,
  and Angell}}]{Q1991}
\bibinfo{author}{\bibfnamefont{Q.}~\bibnamefont{Zheng}},
  \bibinfo{author}{\bibfnamefont{D.}~\bibnamefont{Durben}},
  \bibinfo{author}{\bibfnamefont{G.}~\bibnamefont{Wolf}}, \bibnamefont{and}
  \bibinfo{author}{\bibfnamefont{C.}~\bibnamefont{Angell}},
  \bibinfo{journal}{Science} \textbf{\bibinfo{volume}{254}},
  \bibinfo{pages}{829} (\bibinfo{year}{1991}).

\bibitem[{\citenamefont{Imre et~al.}(2002)\citenamefont{Imre, Maris, and
  Williams}}]{Imre2002}
\bibinfo{author}{\bibfnamefont{A.~R.} \bibnamefont{Imre}},
  \bibinfo{author}{\bibfnamefont{H.~J.} \bibnamefont{Maris}}, \bibnamefont{and}
  \bibinfo{author}{\bibfnamefont{P.~R.} \bibnamefont{Williams}},
  \emph{\bibinfo{title}{{Liquids Under Negative Pressure : Proceedings of the
  NATO Advanced Research Workshop of Liquids Under Negative Pressure Budapest,
  Hungary 23-25 February 2002}}} (\bibinfo{publisher}{Springer Netherlands},
  \bibinfo{year}{2002}), ISBN \bibinfo{isbn}{9781402008962}.

\bibitem[{\citenamefont{Caupin}(2015)}]{Caupin2015}
\bibinfo{author}{\bibfnamefont{F.}~\bibnamefont{Caupin}},
  \bibinfo{journal}{Journal of Non-Crystalline Solids}
  \textbf{\bibinfo{volume}{407}}, \bibinfo{pages}{441} (\bibinfo{year}{2015}).

\bibitem[{\citenamefont{Gallo et~al.}(2016)\citenamefont{Gallo, Amann-Winkel,
  Angell, Anisimov, Caupin, Chakravarty, Lascaris, Loerting, Panagiotopoulos,
  Russo et~al.}}]{gallo2016}
\bibinfo{author}{\bibfnamefont{P.}~\bibnamefont{Gallo}},
  \bibinfo{author}{\bibfnamefont{K.}~\bibnamefont{Amann-Winkel}},
  \bibinfo{author}{\bibfnamefont{C.~A.} \bibnamefont{Angell}},
  \bibinfo{author}{\bibfnamefont{M.~A.} \bibnamefont{Anisimov}},
  \bibinfo{author}{\bibfnamefont{F.}~\bibnamefont{Caupin}},
  \bibinfo{author}{\bibfnamefont{C.}~\bibnamefont{Chakravarty}},
  \bibinfo{author}{\bibfnamefont{E.}~\bibnamefont{Lascaris}},
  \bibinfo{author}{\bibfnamefont{T.}~\bibnamefont{Loerting}},
  \bibinfo{author}{\bibfnamefont{A.~Z.} \bibnamefont{Panagiotopoulos}},
  \bibinfo{author}{\bibfnamefont{J.}~\bibnamefont{Russo}},
  \bibnamefont{et~al.}, \bibinfo{journal}{Chemical Reviews}
  \textbf{\bibinfo{volume}{116}}, \bibinfo{pages}{7463} (\bibinfo{year}{2016}).

\bibitem[{\citenamefont{Montero~de Hijes et~al.}(2018)\citenamefont{Montero~de
  Hijes, Sanz, Joly, Valeriani, and Caupin}}]{pablo_D}
\bibinfo{author}{\bibfnamefont{P.}~\bibnamefont{Montero~de Hijes}},
  \bibinfo{author}{\bibfnamefont{E.}~\bibnamefont{Sanz}},
  \bibinfo{author}{\bibfnamefont{L.}~\bibnamefont{Joly}},
  \bibinfo{author}{\bibfnamefont{C.}~\bibnamefont{Valeriani}},
  \bibnamefont{and} \bibinfo{author}{\bibfnamefont{F.}~\bibnamefont{Caupin}},
  \bibinfo{journal}{J. Chem. Phys.} \textbf{\bibinfo{volume}{149}},
  \bibinfo{pages}{094503} (\bibinfo{year}{2018}).

\bibitem[{\citenamefont{Marcolli}(2017)}]{Marcolli2017}
\bibinfo{author}{\bibfnamefont{C.}~\bibnamefont{Marcolli}},
  \bibinfo{journal}{Scientific Reports} \textbf{\bibinfo{volume}{7}},
  \bibinfo{pages}{16634} (\bibinfo{year}{2017}).

\bibitem[{\citenamefont{Kanno et~al.}(1975)\citenamefont{Kanno, Speedy, and
  Angell}}]{KannoScience1975}
\bibinfo{author}{\bibfnamefont{H.}~\bibnamefont{Kanno}},
  \bibinfo{author}{\bibfnamefont{R.~J.} \bibnamefont{Speedy}},
  \bibnamefont{and} \bibinfo{author}{\bibfnamefont{C.~A.}
  \bibnamefont{Angell}}, \bibinfo{journal}{Science}
  \textbf{\bibinfo{volume}{189}}, \bibinfo{pages}{880} (\bibinfo{year}{1975}).

\bibitem[{\citenamefont{Abascal et~al.}(2005)\citenamefont{Abascal, Sanz,
  {Garc{\'{i}}a Fern{\'{a}}ndez}, and Vega}}]{Abascal2005}
\bibinfo{author}{\bibfnamefont{J.~L.~F.} \bibnamefont{Abascal}},
  \bibinfo{author}{\bibfnamefont{E.}~\bibnamefont{Sanz}},
  \bibinfo{author}{\bibfnamefont{R.}~\bibnamefont{{Garc{\'{i}}a
  Fern{\'{a}}ndez}}}, \bibnamefont{and}
  \bibinfo{author}{\bibfnamefont{C.}~\bibnamefont{Vega}}, \bibinfo{journal}{The
  Journal of Chemical Physics} \textbf{\bibinfo{volume}{122}},
  \bibinfo{pages}{234511} (\bibinfo{year}{2005}).

\bibitem[{\citenamefont{Debenedetti et~al.}(2020)\citenamefont{Debenedetti,
  Sciortino, and Zerze}}]{science_2020}
\bibinfo{author}{\bibfnamefont{P.~G.} \bibnamefont{Debenedetti}},
  \bibinfo{author}{\bibfnamefont{F.}~\bibnamefont{Sciortino}},
  \bibnamefont{and} \bibinfo{author}{\bibfnamefont{H.}~\bibnamefont{Zerze}},
  \bibinfo{journal}{Science} \textbf{\bibinfo{volume}{369}},
  \bibinfo{pages}{289} (\bibinfo{year}{2020}).

\bibitem[{\citenamefont{Hess et~al.}(2008)\citenamefont{Hess, Kutzner, van~der
  Spoel, and Lindahl}}]{Hess2008}
\bibinfo{author}{\bibfnamefont{B.}~\bibnamefont{Hess}},
  \bibinfo{author}{\bibfnamefont{C.}~\bibnamefont{Kutzner}},
  \bibinfo{author}{\bibfnamefont{D.}~\bibnamefont{van~der Spoel}},
  \bibnamefont{and} \bibinfo{author}{\bibfnamefont{E.}~\bibnamefont{Lindahl}},
  \textbf{\bibinfo{volume}{4}}, \bibinfo{pages}{435} (\bibinfo{year}{2008}).

\bibitem[{\citenamefont{Matsui et~al.}(2019)\citenamefont{Matsui, Yagasaki,
  Matsumoto, and Tanaka}}]{Matsui}
\bibinfo{author}{\bibfnamefont{T.}~\bibnamefont{Matsui}},
  \bibinfo{author}{\bibfnamefont{T.}~\bibnamefont{Yagasaki}},
  \bibinfo{author}{\bibfnamefont{M.}~\bibnamefont{Matsumoto}},
  \bibnamefont{and} \bibinfo{author}{\bibfnamefont{H.}~\bibnamefont{Tanaka}},
  \bibinfo{journal}{The Journal of chemical physics}
  \textbf{\bibinfo{volume}{150}}, \bibinfo{pages}{041102}
  (\bibinfo{year}{2019}).

\bibitem[{\citenamefont{Falenty et~al.}(2014)\citenamefont{Falenty, Hansen, and
  Kuhs}}]{icexvi}
\bibinfo{author}{\bibfnamefont{A.}~\bibnamefont{Falenty}},
  \bibinfo{author}{\bibfnamefont{T.~C.} \bibnamefont{Hansen}},
  \bibnamefont{and} \bibinfo{author}{\bibfnamefont{W.~F.} \bibnamefont{Kuhs}},
  \bibinfo{journal}{Nature} \textbf{\bibinfo{volume}{516}},
  \bibinfo{pages}{231} (\bibinfo{year}{2014}).

\bibitem[{SM_()}]{SM_ref}
\emph{\bibinfo{title}{See Supplemental Material at http://link.aps.org/
  supplemental/10.1103/PhysRevLett.126.015704 for a detailed description of the
  numerical methods and results (seeding, direct coexistence, mislabeling,
  thermodynamic integration and attachment rates) of this work and the fitting
  curves shown in the main text.}} (????).

\bibitem[{\citenamefont{Conde et~al.}(2009)\citenamefont{Conde, Vega, Tribello,
  and Slater}}]{virtual}
\bibinfo{author}{\bibfnamefont{M.}~\bibnamefont{Conde}},
  \bibinfo{author}{\bibfnamefont{C.}~\bibnamefont{Vega}},
  \bibinfo{author}{\bibfnamefont{G.}~\bibnamefont{Tribello}}, \bibnamefont{and}
  \bibinfo{author}{\bibfnamefont{B.}~\bibnamefont{Slater}},
  \bibinfo{journal}{The Journal of chemical physics}
  \textbf{\bibinfo{volume}{131}}, \bibinfo{pages}{034510}
  (\bibinfo{year}{2009}).

\bibitem[{\citenamefont{Speedy}(1982)}]{Speedy1982}
\bibinfo{author}{\bibfnamefont{R.~J.} \bibnamefont{Speedy}},
  \bibinfo{journal}{The Journal of Physical Chemistry}
  \textbf{\bibinfo{volume}{86}}, \bibinfo{pages}{982} (\bibinfo{year}{1982}).

\bibitem[{\citenamefont{Rovigatti et~al.}(2017)\citenamefont{Rovigatti, Bianco,
  Tavares, and Sciortino}}]{Rovigatti2017}
\bibinfo{author}{\bibfnamefont{L.}~\bibnamefont{Rovigatti}},
  \bibinfo{author}{\bibfnamefont{V.}~\bibnamefont{Bianco}},
  \bibinfo{author}{\bibfnamefont{J.}~\bibnamefont{Tavares}}, \bibnamefont{and}
  \bibinfo{author}{\bibfnamefont{F.}~\bibnamefont{Sciortino}},
  \bibinfo{journal}{Journal of Chemical Physics}
  \textbf{\bibinfo{volume}{146}}, \bibinfo{pages}{041103}
  (\bibinfo{year}{2017}).

\bibitem[{\citenamefont{Debenedetti}(2003)}]{De03}
\bibinfo{author}{\bibfnamefont{P.~G.} \bibnamefont{Debenedetti}},
  \bibinfo{journal}{Journal of Physics: Condensed Matter}
  \textbf{\bibinfo{volume}{15}}, \bibinfo{pages}{R1669} (\bibinfo{year}{2003}).

\bibitem[{\citenamefont{Bridgman}(1912)}]{bridgman}
\bibinfo{author}{\bibfnamefont{P.~W.} \bibnamefont{Bridgman}},
  \textbf{\bibinfo{volume}{47}}, \bibinfo{pages}{441} (\bibinfo{year}{1912}).

\bibitem[{\citenamefont{Speedy and Angell}(1976)}]{speedy76}
\bibinfo{author}{\bibfnamefont{R.~J.} \bibnamefont{Speedy}} \bibnamefont{and}
  \bibinfo{author}{\bibfnamefont{C.~A.} \bibnamefont{Angell}},
  \bibinfo{journal}{J. Chem. Phys.} \textbf{\bibinfo{volume}{65}},
  \bibinfo{pages}{851} (\bibinfo{year}{1976}).

\bibitem[{\citenamefont{Poole et~al.}(1992)\citenamefont{Poole, Sciortino,
  Essmann, and Stanley}}]{llcp}
\bibinfo{author}{\bibfnamefont{P.~H.} \bibnamefont{Poole}},
  \bibinfo{author}{\bibfnamefont{F.}~\bibnamefont{Sciortino}},
  \bibinfo{author}{\bibfnamefont{U.}~\bibnamefont{Essmann}}, \bibnamefont{and}
  \bibinfo{author}{\bibfnamefont{H.~E.} \bibnamefont{Stanley}},
  \bibinfo{journal}{Nature} \textbf{\bibinfo{volume}{360}},
  \bibinfo{pages}{324} (\bibinfo{year}{1992}).

\bibitem[{\citenamefont{Kim et~al.}(2009)\citenamefont{Kim, Barstow, Tate, and
  Gruner}}]{Kim:2009aa}
\bibinfo{author}{\bibfnamefont{C.~U.} \bibnamefont{Kim}},
  \bibinfo{author}{\bibfnamefont{B.}~\bibnamefont{Barstow}},
  \bibinfo{author}{\bibfnamefont{M.~W.} \bibnamefont{Tate}}, \bibnamefont{and}
  \bibinfo{author}{\bibfnamefont{S.~M.} \bibnamefont{Gruner}},
  \bibinfo{journal}{Proceedings of the National Academy of Sciences}
  \textbf{\bibinfo{volume}{106}}, \bibinfo{pages}{4596} (\bibinfo{year}{2009}).

\bibitem[{\citenamefont{Stokely et~al.}(2010)\citenamefont{Stokely, Mazza,
  Stanley, and Franzese}}]{Stokely2010}
\bibinfo{author}{\bibfnamefont{K.}~\bibnamefont{Stokely}},
  \bibinfo{author}{\bibfnamefont{M.~G.} \bibnamefont{Mazza}},
  \bibinfo{author}{\bibfnamefont{H.~E.} \bibnamefont{Stanley}},
  \bibnamefont{and} \bibinfo{author}{\bibfnamefont{G.}~\bibnamefont{Franzese}},
  \bibinfo{journal}{Proceedings of the National Academy of Sciences of the
  United States of America} \textbf{\bibinfo{volume}{107}},
  \bibinfo{pages}{1301} (\bibinfo{year}{2010}).

\bibitem[{\citenamefont{Bianco and Franzese}(2014)}]{Bianco2014}
\bibinfo{author}{\bibfnamefont{V.}~\bibnamefont{Bianco}} \bibnamefont{and}
  \bibinfo{author}{\bibfnamefont{G.}~\bibnamefont{Franzese}},
  \bibinfo{journal}{Scientific reports} \textbf{\bibinfo{volume}{4}},
  \bibinfo{pages}{4440} (\bibinfo{year}{2014}).

\bibitem[{\citenamefont{Palmer et~al.}(2014)\citenamefont{Palmer, Martelli,
  Liu, Car, Panagiotopoulos, and Debenedetti}}]{Palmer2014}
\bibinfo{author}{\bibfnamefont{J.~C.} \bibnamefont{Palmer}},
  \bibinfo{author}{\bibfnamefont{F.}~\bibnamefont{Martelli}},
  \bibinfo{author}{\bibfnamefont{Y.}~\bibnamefont{Liu}},
  \bibinfo{author}{\bibfnamefont{R.}~\bibnamefont{Car}},
  \bibinfo{author}{\bibfnamefont{A.~Z.} \bibnamefont{Panagiotopoulos}},
  \bibnamefont{and} \bibinfo{author}{\bibfnamefont{P.~G.}
  \bibnamefont{Debenedetti}}, \bibinfo{journal}{Nature}
  \textbf{\bibinfo{volume}{510}}, \bibinfo{pages}{385} (\bibinfo{year}{2014}).

\bibitem[{\citenamefont{Pallares et~al.}(2014)\citenamefont{Pallares, {El Mekki
  Azouzi}, Gonz{\'{a}}lez, Aragones, Abascal, Valeriani, and
  Caupin}}]{Pallares2014}
\bibinfo{author}{\bibfnamefont{G.}~\bibnamefont{Pallares}},
  \bibinfo{author}{\bibfnamefont{M.}~\bibnamefont{{El Mekki Azouzi}}},
  \bibinfo{author}{\bibfnamefont{M.~A.} \bibnamefont{Gonz{\'{a}}lez}},
  \bibinfo{author}{\bibfnamefont{J.~L.} \bibnamefont{Aragones}},
  \bibinfo{author}{\bibfnamefont{J.~F.} \bibnamefont{Abascal}},
  \bibinfo{author}{\bibfnamefont{C.}~\bibnamefont{Valeriani}},
  \bibnamefont{and} \bibinfo{author}{\bibfnamefont{F.}~\bibnamefont{Caupin}},
  \bibinfo{journal}{Proceedings of the National Academy of Sciences}
  \textbf{\bibinfo{volume}{111}}, \bibinfo{pages}{7936} (\bibinfo{year}{2014}).

\bibitem[{\citenamefont{Nilsson and Pettersson}(2015)}]{nilsson_2015}
\bibinfo{author}{\bibfnamefont{A.}~\bibnamefont{Nilsson}} \bibnamefont{and}
  \bibinfo{author}{\bibfnamefont{L.~G.} \bibnamefont{Pettersson}},
  \bibinfo{journal}{Nature Communications} \textbf{\bibinfo{volume}{6}},
  \bibinfo{pages}{8998} (\bibinfo{year}{2015}).

\bibitem[{\citenamefont{Perakis et~al.}(2017)\citenamefont{Perakis,
  Amann-Winkel, Lehmk{\"{u}}hler, Sprung, Mariedahl, Sellberg, Pathak,
  Sp{\"{a}}h, Cavalca, Schlesinger et~al.}}]{Perakis2017}
\bibinfo{author}{\bibfnamefont{F.}~\bibnamefont{Perakis}},
  \bibinfo{author}{\bibfnamefont{K.}~\bibnamefont{Amann-Winkel}},
  \bibinfo{author}{\bibfnamefont{F.}~\bibnamefont{Lehmk{\"{u}}hler}},
  \bibinfo{author}{\bibfnamefont{M.}~\bibnamefont{Sprung}},
  \bibinfo{author}{\bibfnamefont{D.}~\bibnamefont{Mariedahl}},
  \bibinfo{author}{\bibfnamefont{J.~A.} \bibnamefont{Sellberg}},
  \bibinfo{author}{\bibfnamefont{H.}~\bibnamefont{Pathak}},
  \bibinfo{author}{\bibfnamefont{A.}~\bibnamefont{Sp{\"{a}}h}},
  \bibinfo{author}{\bibfnamefont{F.}~\bibnamefont{Cavalca}},
  \bibinfo{author}{\bibfnamefont{D.}~\bibnamefont{Schlesinger}},
  \bibnamefont{et~al.}, \bibinfo{journal}{Proceedings of the National Academy
  of Sciences} \textbf{\bibinfo{volume}{114}}, \bibinfo{pages}{8193}
  (\bibinfo{year}{2017}).

\bibitem[{\citenamefont{Bianco and Franzese}(2019)}]{Bianco2019}
\bibinfo{author}{\bibfnamefont{V.}~\bibnamefont{Bianco}} \bibnamefont{and}
  \bibinfo{author}{\bibfnamefont{G.}~\bibnamefont{Franzese}},
  \bibinfo{journal}{Journal of Molecular Liquids}
  \textbf{\bibinfo{volume}{285}}, \bibinfo{pages}{727} (\bibinfo{year}{2019}).

\bibitem[{\citenamefont{Gallo et~al.}(2019)\citenamefont{Gallo, Loerting, and
  Sciortino}}]{Gallo2019}
\bibinfo{author}{\bibfnamefont{P.}~\bibnamefont{Gallo}},
  \bibinfo{author}{\bibfnamefont{T.}~\bibnamefont{Loerting}}, \bibnamefont{and}
  \bibinfo{author}{\bibfnamefont{F.}~\bibnamefont{Sciortino}},
  \bibinfo{journal}{The Journal of Chemical Physics}
  \textbf{\bibinfo{volume}{151}}, \bibinfo{pages}{210401}
  (\bibinfo{year}{2019}).

\bibitem[{\citenamefont{Bai and Li}(2005)}]{Bai2005}
\bibinfo{author}{\bibfnamefont{X.-M.} \bibnamefont{Bai}} \bibnamefont{and}
  \bibinfo{author}{\bibfnamefont{M.}~\bibnamefont{Li}}, \bibinfo{journal}{The
  Journal of Chemical Physics} \textbf{\bibinfo{volume}{122}},
  \bibinfo{pages}{224510} (\bibinfo{year}{2005}).

\bibitem[{\citenamefont{Espinosa et~al.}(2014)\citenamefont{Espinosa, Sanz,
  Valeriani, and Vega}}]{Espinosa2014}
\bibinfo{author}{\bibfnamefont{J.~R.} \bibnamefont{Espinosa}},
  \bibinfo{author}{\bibfnamefont{E.}~\bibnamefont{Sanz}},
  \bibinfo{author}{\bibfnamefont{C.}~\bibnamefont{Valeriani}},
  \bibnamefont{and} \bibinfo{author}{\bibfnamefont{C.}~\bibnamefont{Vega}},
  \bibinfo{journal}{The Journal of chemical physics}
  \textbf{\bibinfo{volume}{141}}, \bibinfo{pages}{18C529}
  (\bibinfo{year}{2014}).

\bibitem[{\citenamefont{Espinosa
  et~al.}(2016{\natexlab{b}})\citenamefont{Espinosa, Vega, Valeriani, and
  Sanz}}]{Espinosa2016a}
\bibinfo{author}{\bibfnamefont{J.~R.} \bibnamefont{Espinosa}},
  \bibinfo{author}{\bibfnamefont{C.}~\bibnamefont{Vega}},
  \bibinfo{author}{\bibfnamefont{C.}~\bibnamefont{Valeriani}},
  \bibnamefont{and} \bibinfo{author}{\bibfnamefont{E.}~\bibnamefont{Sanz}},
  \bibinfo{journal}{The Journal of Chemical Physics}
  \textbf{\bibinfo{volume}{144}}, \bibinfo{pages}{034501}
  (\bibinfo{year}{2016}{\natexlab{b}}).

\bibitem[{\citenamefont{Espinosa
  et~al.}(2016{\natexlab{c}})\citenamefont{Espinosa, Navarro, Sanz, Valeriani,
  and Vega}}]{Espinosa2016b}
\bibinfo{author}{\bibfnamefont{J.~R.} \bibnamefont{Espinosa}},
  \bibinfo{author}{\bibfnamefont{C.}~\bibnamefont{Navarro}},
  \bibinfo{author}{\bibfnamefont{E.}~\bibnamefont{Sanz}},
  \bibinfo{author}{\bibfnamefont{C.}~\bibnamefont{Valeriani}},
  \bibnamefont{and} \bibinfo{author}{\bibfnamefont{C.}~\bibnamefont{Vega}},
  \bibinfo{journal}{The Journal of Chemical Physics}
  \textbf{\bibinfo{volume}{145}}, \bibinfo{pages}{211922}
  (\bibinfo{year}{2016}{\natexlab{c}}).

\bibitem[{\citenamefont{Espinosa et~al.}(2018)\citenamefont{Espinosa, Vega, and
  Sanz}}]{Espinosa2018}
\bibinfo{author}{\bibfnamefont{J.~R.} \bibnamefont{Espinosa}},
  \bibinfo{author}{\bibfnamefont{C.}~\bibnamefont{Vega}}, \bibnamefont{and}
  \bibinfo{author}{\bibfnamefont{E.}~\bibnamefont{Sanz}}, \bibinfo{journal}{The
  Journal of Physical Chemistry C} \textbf{\bibinfo{volume}{122}},
  \bibinfo{pages}{22892} (\bibinfo{year}{2018}).

\bibitem[{\citenamefont{Lifanov et~al.}(2016)\citenamefont{Lifanov, Vorselaars,
  and Quigley}}]{Lifanov2016}
\bibinfo{author}{\bibfnamefont{Y.}~\bibnamefont{Lifanov}},
  \bibinfo{author}{\bibfnamefont{B.}~\bibnamefont{Vorselaars}},
  \bibnamefont{and} \bibinfo{author}{\bibfnamefont{D.}~\bibnamefont{Quigley}},
  \bibinfo{journal}{The Journal of Chemical Physics}
  \textbf{\bibinfo{volume}{145}}, \bibinfo{pages}{211912}
  (\bibinfo{year}{2016}).

\bibitem[{\citenamefont{Leoni et~al.}(2019)\citenamefont{Leoni, Shi, Tanaka,
  and Russo}}]{Leoni2019}
\bibinfo{author}{\bibfnamefont{F.}~\bibnamefont{Leoni}},
  \bibinfo{author}{\bibfnamefont{R.}~\bibnamefont{Shi}},
  \bibinfo{author}{\bibfnamefont{H.}~\bibnamefont{Tanaka}}, \bibnamefont{and}
  \bibinfo{author}{\bibfnamefont{J.}~\bibnamefont{Russo}},
  \bibinfo{journal}{The Journal of Chemical Physics}
  \textbf{\bibinfo{volume}{151}}, \bibinfo{pages}{044505}
  (\bibinfo{year}{2019}).

\bibitem[{\citenamefont{Montero~de Hijes et~al.}(2019)\citenamefont{Montero~de
  Hijes, Espinosa, Sanz, and Vega}}]{montero2019}
\bibinfo{author}{\bibfnamefont{P.}~\bibnamefont{Montero~de Hijes}},
  \bibinfo{author}{\bibfnamefont{J.~R.} \bibnamefont{Espinosa}},
  \bibinfo{author}{\bibfnamefont{E.}~\bibnamefont{Sanz}}, \bibnamefont{and}
  \bibinfo{author}{\bibfnamefont{C.}~\bibnamefont{Vega}}, \bibinfo{journal}{The
  Journal of Chemical Physics} \textbf{\bibinfo{volume}{151}},
  \bibinfo{pages}{144501} (\bibinfo{year}{2019}).

\bibitem[{\citenamefont{{Montero de Hijes} et~al.}(2020)\citenamefont{{Montero
  de Hijes}, Espinosa, Bianco, Sanz, and Vega}}]{MonterodeHijes2020}
\bibinfo{author}{\bibfnamefont{P.}~\bibnamefont{{Montero de Hijes}}},
  \bibinfo{author}{\bibfnamefont{J.~R.} \bibnamefont{Espinosa}},
  \bibinfo{author}{\bibfnamefont{V.}~\bibnamefont{Bianco}},
  \bibinfo{author}{\bibfnamefont{E.}~\bibnamefont{Sanz}}, \bibnamefont{and}
  \bibinfo{author}{\bibfnamefont{C.}~\bibnamefont{Vega}}, \bibinfo{journal}{The
  Journal of Physical Chemistry C} \textbf{\bibinfo{volume}{124}},
  \bibinfo{pages}{8795 } (\bibinfo{year}{2020}).

\bibitem[{\citenamefont{Montero~de Hijes et~al.}(2020)\citenamefont{Montero~de
  Hijes, Shi, Noya, Santiso, Gubbins, Sanz, and Vega}}]{montero2020young}
\bibinfo{author}{\bibfnamefont{P.}~\bibnamefont{Montero~de Hijes}},
  \bibinfo{author}{\bibfnamefont{K.}~\bibnamefont{Shi}},
  \bibinfo{author}{\bibfnamefont{E.}~\bibnamefont{Noya}},
  \bibinfo{author}{\bibfnamefont{E.}~\bibnamefont{Santiso}},
  \bibinfo{author}{\bibfnamefont{K.}~\bibnamefont{Gubbins}},
  \bibinfo{author}{\bibfnamefont{E.}~\bibnamefont{Sanz}}, \bibnamefont{and}
  \bibinfo{author}{\bibfnamefont{C.}~\bibnamefont{Vega}}, \bibinfo{journal}{The
  Journal of Chemical Physics} \textbf{\bibinfo{volume}{153}},
  \bibinfo{pages}{191102} (\bibinfo{year}{2020}).

\bibitem[{\citenamefont{Tipeev et~al.}(2018)\citenamefont{Tipeev, Zanotto, and
  Rino}}]{tipeev}
\bibinfo{author}{\bibfnamefont{A.~O.} \bibnamefont{Tipeev}},
  \bibinfo{author}{\bibfnamefont{E.~D.} \bibnamefont{Zanotto}},
  \bibnamefont{and} \bibinfo{author}{\bibfnamefont{J.~P.} \bibnamefont{Rino}},
  \bibinfo{journal}{The Journal of Physical Chemistry C}
  \textbf{\bibinfo{volume}{122}}, \bibinfo{pages}{28884}
  (\bibinfo{year}{2018}).

\bibitem[{\citenamefont{Knott et~al.}(2012)\citenamefont{Knott, Molinero,
  Doherty, and Peters}}]{Knott2012}
\bibinfo{author}{\bibfnamefont{B.~C.} \bibnamefont{Knott}},
  \bibinfo{author}{\bibfnamefont{V.}~\bibnamefont{Molinero}},
  \bibinfo{author}{\bibfnamefont{M.~F.} \bibnamefont{Doherty}},
  \bibnamefont{and} \bibinfo{author}{\bibfnamefont{B.}~\bibnamefont{Peters}},
  \bibinfo{journal}{Journal of the American Chemical Society}
  \textbf{\bibinfo{volume}{134}}, \bibinfo{pages}{19544}
  (\bibinfo{year}{2012}).

\bibitem[{\citenamefont{Pereyra et~al.}(2011)\citenamefont{Pereyra, Szleifer,
  and Carignano}}]{pereyra2011}
\bibinfo{author}{\bibfnamefont{R.~G.} \bibnamefont{Pereyra}},
  \bibinfo{author}{\bibfnamefont{I.}~\bibnamefont{Szleifer}}, \bibnamefont{and}
  \bibinfo{author}{\bibfnamefont{M.~A.} \bibnamefont{Carignano}},
  \bibinfo{journal}{The Journal of chemical physics}
  \textbf{\bibinfo{volume}{135}}, \bibinfo{pages}{034508}
  (\bibinfo{year}{2011}).

\bibitem[{\citenamefont{Dasgupta et~al.}(2020)\citenamefont{Dasgupta, Coli, and
  Dijkstra}}]{Dasgupta2020}
\bibinfo{author}{\bibfnamefont{T.}~\bibnamefont{Dasgupta}},
  \bibinfo{author}{\bibfnamefont{G.~M.} \bibnamefont{Coli}}, \bibnamefont{and}
  \bibinfo{author}{\bibfnamefont{M.}~\bibnamefont{Dijkstra}},
  \bibinfo{journal}{ACS Nano} \textbf{\bibinfo{volume}{14}},
  \bibinfo{pages}{3957} (\bibinfo{year}{2020}).

\bibitem[{\citenamefont{Niu et~al.}(2020)\citenamefont{Niu, Bonati, Piaggi, and
  Parrinello}}]{Niu2020}
\bibinfo{author}{\bibfnamefont{H.}~\bibnamefont{Niu}},
  \bibinfo{author}{\bibfnamefont{L.}~\bibnamefont{Bonati}},
  \bibinfo{author}{\bibfnamefont{P.~M.} \bibnamefont{Piaggi}},
  \bibnamefont{and}
  \bibinfo{author}{\bibfnamefont{M.}~\bibnamefont{Parrinello}},
  \bibinfo{journal}{Nature Communications} \textbf{\bibinfo{volume}{11}},
  \bibinfo{pages}{2654} (\bibinfo{year}{2020}).

\bibitem[{\citenamefont{Zimmermann et~al.}(2015)\citenamefont{Zimmermann,
  Vorselaars, Quigley, and Peters}}]{zimmermann2015}
\bibinfo{author}{\bibfnamefont{N.~E.} \bibnamefont{Zimmermann}},
  \bibinfo{author}{\bibfnamefont{B.}~\bibnamefont{Vorselaars}},
  \bibinfo{author}{\bibfnamefont{D.}~\bibnamefont{Quigley}}, \bibnamefont{and}
  \bibinfo{author}{\bibfnamefont{B.}~\bibnamefont{Peters}},
  \bibinfo{journal}{Journal of the American Chemical Society}
  \textbf{\bibinfo{volume}{137}}, \bibinfo{pages}{13352}
  (\bibinfo{year}{2015}).

\bibitem[{\citenamefont{Lechner and Dellago}(2008)}]{Lechner2008}
\bibinfo{author}{\bibfnamefont{W.}~\bibnamefont{Lechner}} \bibnamefont{and}
  \bibinfo{author}{\bibfnamefont{C.}~\bibnamefont{Dellago}},
  \bibinfo{journal}{The Journal of Chemical Physics}
  \textbf{\bibinfo{volume}{129}}, \bibinfo{pages}{114707}
  (\bibinfo{year}{2008}).

\bibitem[{\citenamefont{Becker and D{\"{o}}ring}(1935)}]{Becker1935}
\bibinfo{author}{\bibfnamefont{R.}~\bibnamefont{Becker}} \bibnamefont{and}
  \bibinfo{author}{\bibfnamefont{W.}~\bibnamefont{D{\"{o}}ring}},
  \bibinfo{journal}{Annalen der Physik} \textbf{\bibinfo{volume}{416}},
  \bibinfo{pages}{719} (\bibinfo{year}{1935}).

\bibitem[{\citenamefont{Auer and Frenkel}(2001)}]{auer2001}
\bibinfo{author}{\bibfnamefont{S.}~\bibnamefont{Auer}} \bibnamefont{and}
  \bibinfo{author}{\bibfnamefont{D.}~\bibnamefont{Frenkel}},
  \bibinfo{journal}{Nature} \textbf{\bibinfo{volume}{409}},
  \bibinfo{pages}{1020} (\bibinfo{year}{2001}).

\bibitem[{\citenamefont{Altabet et~al.}(2017)\citenamefont{Altabet, Singh,
  Stillinger, and Debenedetti}}]{altabet2017}
\bibinfo{author}{\bibfnamefont{Y.~E.} \bibnamefont{Altabet}},
  \bibinfo{author}{\bibfnamefont{R.~S.} \bibnamefont{Singh}},
  \bibinfo{author}{\bibfnamefont{F.~H.} \bibnamefont{Stillinger}},
  \bibnamefont{and} \bibinfo{author}{\bibfnamefont{P.~G.}
  \bibnamefont{Debenedetti}}, \bibinfo{journal}{Langmuir}
  \textbf{\bibinfo{volume}{33}}, \bibinfo{pages}{11771} (\bibinfo{year}{2017}).

\bibitem[{\citenamefont{Holten et~al.}(2017)\citenamefont{Holten, Qiu,
  Guillerm, Wilke, Ricka, Frenz, and Caupin}}]{holten2017}
\bibinfo{author}{\bibfnamefont{V.}~\bibnamefont{Holten}},
  \bibinfo{author}{\bibfnamefont{C.}~\bibnamefont{Qiu}},
  \bibinfo{author}{\bibfnamefont{E.}~\bibnamefont{Guillerm}},
  \bibinfo{author}{\bibfnamefont{M.}~\bibnamefont{Wilke}},
  \bibinfo{author}{\bibfnamefont{J.}~\bibnamefont{Ricka}},
  \bibinfo{author}{\bibfnamefont{M.}~\bibnamefont{Frenz}}, \bibnamefont{and}
  \bibinfo{author}{\bibfnamefont{F.}~\bibnamefont{Caupin}},
  \bibinfo{journal}{The journal of physical chemistry letters}
  \textbf{\bibinfo{volume}{8}}, \bibinfo{pages}{5519} (\bibinfo{year}{2017}).

\end{thebibliography}
\end{document}